\documentclass[bibyear]{aa} 
\usepackage{graphicx}
\usepackage{txfonts}
\usepackage{cite}
\usepackage{amsmath}	
\usepackage{amssymb}	
\usepackage{epsfig}
\usepackage{color}
\usepackage{longtable,lscape}
\usepackage{natbib}
\usepackage{graphicx}	
\newcommand{\HI}{\mbox{H\,{\sc i}}}

\newcommand{\kms}{\mbox{km\,s$^{-1}$}}

\newcommand{\Jykms}{\mbox{Jy\,km\,s$^{-1}$}}

\newcommand{\kmsMpc}{\mbox{ km~s$^{-1}$~Mpc$^{-1}$}}
\newcommand{\msun}{$M_\odot$}

\newcommand{\vopt}{\mbox{${v}_{\rm{opt}}$}}

\def\approxlt{\lower.2em\hbox{$\buildrel < \over \sim$}}  
\def\approxgt{\lower.2em\hbox{$\buildrel > \over \sim$}}  

\newcommand{\nan}{Nan\c{c}ay}
\newcommand{\nrt}{NRT}

\newcommand{\eg}{{e.g.}\ }         
\newcommand{\ie}{{i.e.}\ }

\definecolor{grey}{rgb}{0.5,0.6,0.7}

\begin{document}

\title{EZOA: \nan\ H\,{\sc i} follow-up observations in the Zone of Avoidance}

\author{A.C. Schr\"oder \inst{\ref{inst1}}
\and W. van Driel\inst{\ref{inst2} \and \ref{inst3}}
\and R.C. Kraan-Korteweg\inst{\ref{inst4}}
\and C. Belleval\inst{\ref{inst2} \and \ref{inst5}}
          }

\institute{Max-Planck-Institut f\"ur extraterrestrische Physik, Gie{\ss}enbachstra{\ss}e 1, 85748 Garching, Germany, \email{anja@mpe.mpg.de} \label{inst1}
         \and
GEPI, Observatoire de Paris, Universit\'e PSL, CNRS, 5 place Jules Janssen, 92190 Meudon, France \label{inst2}
         \and
Observatoire Radioastronomique de \nan, Observatoire de Paris, Universit\'e PSL, Universit\'e d'Orl\'eans, 18330 \nan, France \label{inst3}
         \and
Department of Astronomy, University of Cape Town, Private Bag X3, Rondebosch 7701, South Africa \label{inst4}
         \and
Universit\'e Paris-Cit\'e, 143 avenue de Versailles, 75016 Paris, France  \label{inst5}
%
             }

   \date{Received ...; accepted ...}

 
  \abstract
   {
We present follow-up 21cm \HI\ line observations made with the \nan\ Radio
Telescope (NRT) of 99 weak or potential detections of galaxies from the
EZOA catalogue in the northern Zone of Avoidance (ZoA), which were
extracted from the shallow version of the EBHIS blind \HI\ survey performed
with the Effelsberg radio telescope. The new NRT observations are on
average almost three times as sensitive as those from EBHIS. Of the 99
observed sources, we confirmed 72, while three yielded inconclusive
results. We find that the quality assessment of the EZOA catalogue entries
correlates well with the NRT recovery rate; for instance, only four of the
22 potential detections could be confirmed. Due to the higher sensitivity
as well as the large north-south extent of the NRT beam, the NRT
observations also yielded five serendipitous detections, which we include
here. We updated the EZOA catalogue with the improved \HI\ parameters and
detections. To test mitigation of Radio Frequency Interference signals, we
also observed selected sources using a dedicated receiver and data
processing system.
}


\maketitle

\nolinenumbers

\section{Introduction}   \label{intro}  

Our current understanding of the local Universe -- such as large scale
structures and the dynamics related to their gravitational pull -- is still
incomplete due to the difficulties in finding galaxies near the obscured
Galactic plane, the so-called Zone of Avoidance (ZoA). Even when
low-latitude galaxies are identified in the optical or near-infrared (NIR),
obtaining reliable redshifts from an optical spectrum remains extremely
difficult \citep{macri19}.

The ZoA is known to obscure and bisect major parts of dynamically important
structures such as the Great Attractor (GA; \citealt{dressler87},
\citealt{kraan96}, and \citealt{woudt04}), the Perseus-Pisces supercluster
(PPS; \citealt{focardi84} and \citealt{chamaraux90}), and the Local Void
(LV; \citealt{tully08} and \citealt{tully19}). Furthermore, newly
identified low-latitude structures are still being discovered to date, such
as the massive Vela supercluster \citep[VSCL;][]{kraan17} at $cz \sim
18,500\,\kms$, which may contribute a significant component to the
(residual) bulk flow motion (\eg \citealt{hudson04} and
\citealt{springob16}).

Observations of the 21\,cm spectral line of neutral hydrogen (\HI) of
galaxies is one of the most successful methods to penetrate the obscuring
dust layers and crowded star fields along the plane of our Galaxy. Their
\HI\ line profiles furthermore provide an immediate measure of their
redshift as well as insight into certain properties such as their total
\HI\ mass. \HI\ surveys therefore are essential towards obtaining a
comprehensive and complete three-dimensional picture of the large-scale
structures in the local Universe, unaffected by absorption of the optical
and NIR emission of galaxies.

For these reasons, numerous systematic large-area \HI\ surveys have been
undertaken, mostly with single dish radio telescopes, but also, more
recently, with interferometers \citep[\eg][]{ramatsoku16}. Pointed
\HI\ observations of optical or NIR-selected ZoA galaxy samples are
complementary to the 'blind' surveys and remain important. Blind surveys
are often quite shallow, and many galaxies can go undetected if their
\HI\ mass or signal-to-noise ratios are low -- the latter especially for
galaxies with broad line widths. Furthermore, cross-correlating the
\HI\ emission with an optical or NIR counterpart can be used to determine
peculiar velocities using the Tully-Fisher relation.

Pointed \HI\ observations of optically detected ZoA galaxies and galaxies
in the 2MASS bright ZoA catalogue \citep[2MZOA][]{schroeder19b} were performed using
both the Parkes 64\,m radio telescope (\eg \citealt{kraan02},
\citealt{schroeder09} and \citealt{said16a}) and the 100\,m class
\nan\ Radio Telescope, NRT, (\citealt{vandriel09} and \citealt{kraan18}).
These studies have proved quite effective in complementing the optically
and NIR-selected galaxy catalogues in the southern and northern ZoA,
respectively. On the other hand, only blind \HI\ surveys can detect
galaxies independent of the dust and star density deep in the plane and
obtain a truly all-sky census of the galaxy distribution within the ZoA.

To date, systematic blind \HI\ surveys have homogeneously covered the
southern hemisphere out to $cz\!\sim\!12,700\,\kms$ with the Multibeam
receiver on the 64\,m Parkes dish, namely HIPASS and its northern extension
($\delta < +25$\degr; \citealt{meyer04} and \citealt{wong06}, respectively)
with a typical rms noise of 13\,mJy\,beam$^{-1}$. In addition, the deeper
Parkes HIZOA survey has an rms noise of 6\,mJy\,beam$^{-1}$ and was
dedicated to the deeply obscured inner part of the ZoA ($|b| < 5\degr$;
\citealt{staveley16} and \citealt{donley05}, respectively). Recently,
significantly deeper \HI\ surveys have become available for the innermost
($|b| \leq 1\fdg5$) southern ZoA. For example, an extraction of \HI\ data
from the SARAO MeerKAT Galactic Plane Legacy Survey
\citep[SMGPS;][]{goedhart24}, for the redshift range $cz < $25,000\,\kms\ and
with a median rms noise level of $\sim 0.45$\,mJy\,beam$^{-1}$, provided
tantalising new insights to the LV \citep{kurapati24}, the Ophiuchus supercluster
\citep{louw24}, the GA region \citep{steyn24a}, and the more distant VSCL
\citep{rajohnson24}.

However, in the northern sky (\ie $\delta \ge +25\fdg5$) no such systematic
blind \HI\ survey existed before 2019. The northern ZoA was surveyed with
the 25\,m Dwingeloo telescope, but at a high rms noise of 40\,mJy
beam$^{-1}$ \citep{henning98,rivers99}, while the HIJASS survey
\citep{lang03}, conducted with the 76\,m Lovell telescope at Jodrell Bank,
only covered parts of the northern sky at a sensitivity comparable to
HIPASS (rms\,$=13-16$\,mJy). Part of the ZoA regions that were accessible
by the 305\,m Arecibo telescope were covered by the ALFAZOA surveys,
comprising the Shallow ALFAZOA survey \citep{sanchez19} covering latitudes
$|b| < 10\degr$ with an rms noise of $5-7$\,mJy beam$^{-1}$, and the Deep
ALFAZOA survey \citep{henning08,mcintyre15} with an rms noise of 1\,mJy
beam$^{-1}$ and covering a much narrower strip in Galactic latitude ($|b|
< 2\degr$). We note that the main ALFALFA survey did not cover the ZoA, see
\citep{alfalfa}.

To complement the systematic southern surveys, the blind Effelsberg--Bonn
\HI\ Survey (EBHIS; \citealt{kerp11} and \citealt{winkel10}) was
launched. It has homogeneously surveyed the entire northern sky ($\delta
\geq -5\degr$) with the 100\,m Effelsberg radio telescope out to $cz =
18\,000\,\kms$ at a nominal rms noise of 16\,mJy, comparable to HIPASS in
the southern sky. For the northern ZoA, \cite{schroeder19} compiled the
so-called EZOA catalogue with 170 detections extracted from the shallower
(first pass) EBHIS survey, which has an rms noise level of
23\,mJy\,beam$^{-1}$ \citep{floer14}.

In this paper, we present higher sensitivity follow-up observations
obtained with the 100\,m class NRT to confirm or improve \HI\ signals of 77
selected EZOA detections and 22 EZOA candidates. We note that compared to
the cube rms level of the EBHIS data, the median rms noise of the detected
EZOA sources is lower (10.4\,mJy\,beam$^{-1}$; \citealt{schroeder19}). For
our follow-up observations, we aimed at reaching a 3\,mJy rms noise level,
similar to our NRT observations of 2MASS bright galaxies \citep{kraan18}).
We also performed Radio Frequency Interference (RFI) mitigation test
observations of a few selected sources, using a dedicated receiver and data
analysis system.

This paper is organised as follows. In Sect.~\ref{sample} the selection of
the observed sources is described, and in Sect.~\ref{observations} the
observations and data reduction. Section~\ref{results} presents the
results, which are discussed in Sect.~\ref{discussion}, finishing with a
summary in Sect.~\ref{summary}. Notes to selected sources are presented in
Appendix~\ref{appnotes}, and the Tables and \HI\ spectra in
Appendix~\ref{apponline}.

Throughout the paper, we use a Hubble constant of $H_{\rm 0}=75$\,\kmsMpc.
For convenience, we introduce the angular distance parameter $b$, which is
expressed in units of the radius of the elliptically-shaped NRT
beam. The beam radius is defined as the angular distance from the pointing
centre at which the telescope gain has diminished to a level of 50\% of its
peak level (\ie 0.5 half-power beam-width, HPBW).

\section{Sample selection }  \label{sample}  

In the pre-release EBHIS data cubes (made from the first of the two
passes), we visually searched the Galactic plane region ($|b| < 6^\circ$)
and extracted 170 \HI\ detections of varying quality, a third of which had
not been previously detected in \HI\ \citep[for details see][]
{schroeder19}.

To select EZOA sources for follow-up \HI\ observations with the NRT we used
the following criteria as a guide line, which resulted in a sample of 77
target sources:
\begin{enumerate}

\item Signal too weak for an accurate determination of the \HI\ profile
  parameters and sky position (peak signal-to-noise ratio (SNR)
  $\lessapprox 10$);

\item No previously published \HI\ detection;

\item \HI\ profiles affected by baseline variations, noise or RFI;

\item Multiple optical or NIR counterparts or uncertain counterpart.

\end{enumerate}

In addition, we selected 22 EZOA candidate sources that were not published
in the EZOA paper. Based on their appearance in the data cube, final
integrated line profile, and existence -- or absence -- of a reasonable
stellar counterpart, they were judged to have a low likelihood of being
real. The NRT follow-up observations of these will, on the one hand, serve
to verify if the emission comes indeed from a galaxy and, on the other
hand, allow an assessment of our internal classification scheme.

Our final tally consists of 99 sources selected for follow-up observations.
Multiple pointings per EZOA source were required for multiple potential
counterparts of an EZOA source or, if no multi-wavelength cross-match
could be identified, to cover the the positional uncertainty area of the
EZOA detection ($r=3\farcm5$) with the the narrower NRT beam (with an east-west
HPBW radius of $1\farcm8$; usually three pointings sufficed). This resulted
in 143 NRT pointings for the 99 sources.

We also selected a few EZOA sources that fall within velocity ranges that
are known to be polluted by strong RFI to apply and test the efficacy of
our RFI mitigation software \citep{belleval19}.

\section{Observations and data reduction}   \label{observations}  

The NRT is a 100 metre-class meridian transit-type instrument (see, \eg
\citealt{vandriel16} for further details on the instrument and data
reduction). Due to the east-west elongated shape of the mirrors, beam size
and sensitivity depend on the observed declination. Its HPBW is always
$3\farcm6$ in right ascension, independent of declination, whereas in
declination it increases from 22$'$ for $\delta<20^{\circ}$ to 31$'$ at
$\delta = 66^{\circ}$, the northern limit of our survey \citep[see
  also][]{matthews00}. Its sensitivity follows the same geometric effect
and decreases correspondingly with declination. The typical system
temperature is 35~K at $\delta = 0\degr$.

Flux calibration is determined through regular measurements of a cold load
calibrator and periodic monitoring of strong continuum sources by the
\nan\ staff. For verification and monitoring of this standard calibration
we also regularly observed \HI\ line flux calibrator galaxies measured at
Arecibo by \citet{oneil04b}. From our 14 NRT measurements we derived a
ratio between our integrated line fluxes and the literature values of
$1.01\pm0.09$.

All sources were observed with the standard auto-correlator (ACRT) using a
set-up of 4096 channels in a 50~MHz bandpass, with a channel spacing of
2.6\,\kms\ and a velocity coverage of $\sim-250$ to 10,600\,\kms. The data
were taken in position-switching mode, with an on-source integration time
of 40 seconds per on-off pair. The observations were made in the period
January 2019\,--\,March 2023, using a total of about 230 hours of telescope
time. 

We used standard NRT software to flag and mitigate strong RFI in the ACRT
data \citep{monnier03c}, average the two receiver polarisations, perform
the declination-dependent conversion from system temperature to flux
density in mJy, fit polynomial baselines, smooth the data to a velocity
resolution of 10\,\kms, and ultimately convert radial velocities to the
optical, heliocentric {\it cz} system. The \HI\ flux was determined
by integrating over the entire velocity width of the detected profile.

Uncertainties in \HI\ line parameters were determined following
\citet{koribalski04}; they depend, among others, on the SNR (ratio of the
line peak flux density and the rms noise level) and the velocity resolution
(10\,\kms ). The integrated \HI\ flux of galaxies of sufficiently large
apparent diameter could in principle be underestimated with the NRT as its
east-west HPBW of $3\farcm6$ is rather small. For example, at the mean
velocity of the detected targets, 3800\,\kms, the NRT HPBW corresponds to a
linear size of 40 kpc.

For the RFI mitigation tests we observed the galaxies simultaneously with
the standard ACRT and with the WIBAR high-resolution and high-sampling rate
receiver. The WIBAR data were processed using the RObust Elusive Line
detection (ROBEL) RFI mitigation software package (see \citealt{belleval19}
for details on WIBAR and ROBEL). For comparison with ACRT data, the
original WIBAR velocity resolution of 0.9\,\kms\ was smoothed to
11.1\,\kms. However, due to receiver problems and the restrictive nature of
the COVID-19 pandemic safety protocols, useful results could only be
obtained for two of the 20 observed test sources.

\section{Results}   \label{results}  

\subsection{Observations} \label{resobs}  

The results of our NRT follow-up observations are summarised in
Table~\ref{tabobs} given in the appendix, including a comparison to
the EBHIS \HI\ measurements. All observations were adjudicated based on the
individual assessment of the spectra (SNR and the profile shape) by three
authors. The resulting NRT spectra are shown in the appendix in
Fig.~\ref{figspectra}.

Listed in Table~\ref{tabobs} are:

Col.\ 1: Identification (ID) of the EZOA source (without the EZOA prefix).

Col.\ 2: Classification of the \HI\ detection. The EBHIS classification is
given to the left of the arrow and the new, NRT, classification to the
right. Class~1 indicates a clear detection; 3+, 3, and $3-$ are detections
with varying degree of confidence, with $3-$ denoting a possible detection
(not listed in the EZOA catalogue); 4 marks a non-detection; and 5
indicates an inconclusive result.

Col.\ 3: Flag $n$ refers to a note in Appendix~\ref{appnotes}.

Col.\ 4: Equatorial coordinates (J2000.0) of the EBHIS detection.

Col.\ 5: EBHIS heliocentric velocity ({\it cz}) and uncertainty, in \kms .

Col.\ 6: EBHIS velocity width at 50\% of peak flux density and associated
uncertainty, in \kms .

Col.\ 7: EBHIS integrated \HI\ flux and associated uncertainty, in Jy\,\kms .

Col.\ 8: EBHIS peak signal-to-noise ratio SNR.

Col.\ 9: NRT pointing ID; a footnote lists the reference of a detection not
observed by us. 

Col.\ 10: NRT pointing position in equatorial (J2000.0) coordinates.

Col.\ 11: NRT pointing target classification: c = cross-match candidate, e
= EBHIS position, s = NRT search position.

Col.\ 12: Distance between the NRT pointing position and the EBHIS source
position, in units of the NRT beam radius $b$.

Col.\ 13: Classification of the cross-match candidate (that is, for
pointing target `c', see Col.\ 11): `d' = definite, `p' = probable, `a' =
ambiguous, `c' = contributor to a confused profile; `u' = uncertain galaxy,
`n' = not a candidate. Where the classification has changed, the former
EBHIS classification is given on the left and the new classification on the
right, linked by an arrow.

Col.\ 14: Class of the NRT/EBHIS integrated flux ratio, $c_{\rm fl}$ (see
Col.\ 15): While $c_{\rm fl}=2$ stands for comparable fluxes (that is, $0.9
\le R_{\rm fl} \le 1.0$; see also the discussion of Fig.~\ref{figcomp}),
classes 1 and 3 indicate a higher and lower NRT flux, respectively; 4 means
no NRT detection, and 5 indicates that the flux ratio is not usable, mainly
when the profiles are not comparable (especially when the line width is
significantly different, \eg due to confusion, or when RFI affects the
profile).

Col.\ 15: Ratio of NRT and EBHIS integrated fluxes, $R_{\rm fl}$.

Col.\ 16: Final NRT flag: 0 = observed candidate is the stellar
counterpart; 1 = pointing with the best detection (no cross-match
candidate); 2 = candidate contributes to the EZOA detection (a plus denotes
a major contributor, a minus a minor contribution); 3 = low-level
detection, not the best pointing; 4 = not detected; 5 = inconclusive (\eg
due to RFI).

Col.\ 17: NRT spectrum rms noise level, in mJy. 

Col.\ 18: NRT heliocentric velocity ({\it cz}) and uncertainty, in \kms .

Col.\ 19: NRT velocity width at 50\% of peak flux density and associated
uncertainty, in \kms .

Col.\ 20: NRT velocity width at 20\% of peak flux density and associated
uncertainty, in \kms .

Col.\ 21: NRT integrated \HI\ flux and associated uncertainty, in Jy\,\kms .

Col.\ 22: NRT peak signal-to-noise ratio SNR.

\subsection{Detections} \label{resdet}  

Based on these observations, we determined the `best' NRT
pointing for a given EZOA source, most of which were towards an optical/NIR cross-match
candidate that is thus confirmed. The decision was made based on the
ratios of fluxes measured towards the various pointings of an EZOA source
(and their spatial separations), profile shapes and possible confusion with
other sources. Where the identification was more complex, we give a note in
Appendix~\ref{appnotes}. In most cases we were able to improve on the
original EZOA SNR. Table~\ref{tabdet} lists the detection of the `best' NRT
pointing, giving the observed and derived parameters as well as information
on the cross-match where one was identified. The columns are as follows.

Col.\ 1: ID of the EZOA source (without the EZOA prefix).

Col.\ 2: Classification of the NRT detection (see Table~\ref{tabobs},
Col.\ 2).

Col.\ 3: Flag $n$ refers to a note in Appendix~\ref{appnotes}.

Col.\ 4: NRT pointing ID (see Table~\ref{tabobs}).

Col.\ 5 -- 9: NRT \HI\ line parameters, from Table~\ref{tabobs}, Cols. 18
-- 22.

Col.\ 10: Radial velocity of the galaxy, in \kms , corrected to the Local
Group frame of reference using
$ v_{\rm LG} = v_{\rm hel} + 300 \sin \ell \cos b $.

Col.\ 11: Distance to the galaxy based on $v_{\rm LG}$, in Mpc.

Col.\ 12: Logarithm of the total \HI\ mass, in \msun .

Col.\ 13: Final NRT flag (see Table~\ref{tabobs}, Col.~16).

Col.\ 14: Final classification of the cross-match (see Table~\ref{tabobs}).

Col.\ 15: Angular distance between NRT pointing and position of the
optical/NIR cross-match (where they differ), in NRT beam radius units,
$b$. Uncertainties in integrated flux and other \HI\ parameters increase
correspondingly. No corrections have been applied to the given parameters.

Col.\ 16: Equatorial (J2000.0) coordinates of the cross-match (where
available) or of the best NRT pointing position.

Col.\ 17: Positional uncertainty: where the visible galaxy appears diffuse,
without a clear centre, we added a colon (:) to indicate a larger
positional uncertainty than implied by the precision of the
coordinates. Where no candidate is found, we give an improved positional
uncertainty ellipse (with minor and major semi-axes in arcminutes).

Col.\ 18 and 19: Galactic longitude and latitude, in degrees.

Col.\ 20: Galactic extinction in the $K$-band, derived from the extinction
maps published by \citeauthor{planck16b} (2016b) using the conversion
$A_K=0.36 \cdot E(B-V)$ and a correction factor $f$ of 0.86
(\citealt{schroeder21}).

Col.\ 21: ID of the cross-match (see Table~\ref{tabgal} for all previously
uncatalogued galaxies around the EZOA sources in our sample).

In addition to the detections of the EZOA sources, we also list
serendipitous detections found in five of the NRT pointings. Their spectra
are shown in Fig.~\ref{fig2spectra} in the Appendix. For three of them we
found a cross-match. 

\subsection{Quality assessment of HI parameters} \label{qa}

\begin{figure}  
\centering
\includegraphics[width=0.45\textwidth]{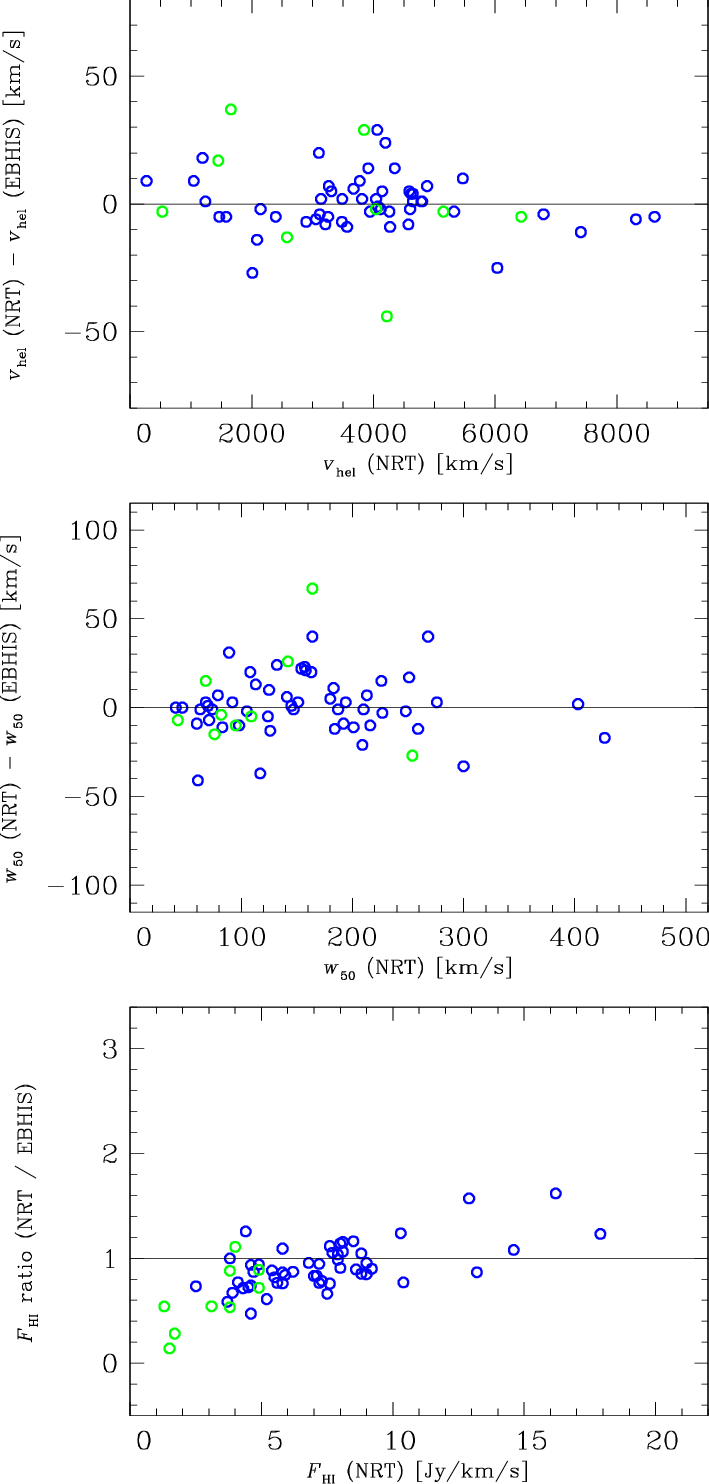}
\caption{Comparison of the EBHIS and NRT \HI\ line parameters central
  velocity (top), the line width (middle), and integrated fluxes
  (bottom). Blue circles are pointings towards cross-matches and green
  circles are the best NRT pointings when no cross-match is visible
  (corresponding to NRT final flag 1).
}
\label{figcomp}
\end{figure}

Figure~\ref{figcomp} shows comparisons of the \HI\ line parameters obtained
with the NRT and EBHIS. We only used non-confused, confirmed EZOA sources
with comparable profiles, resulting in 61 objects (excluding flux ratio
class~5, see Col.~14 in Table~\ref{tabobs}). Pointings with no cross-match
are shown in green (\ie detections with the final NRT flag~1). The
mean NRT$-$EBHIS offsets for the central velocity ($0.6\pm 1.7$\,\kms ) and
line width $w_{\rm 50}$ ($2.0\pm 2.4$\,\kms ) are negligible. The mean flux
ratio of $0.88\pm 0.03$ is slightly low.

\onecolumn
{\scriptsize
\begin{landscape}
\setlength{\tabcolsep}{1.3mm}
\begin{longtable}{
llllr@{\hskip 0.5mm}rr@{\hskip 0.5mm}rr@{\hskip 0.5mm}rr@{\hskip 0.2mm}rrrrrcclllcrrrl
}
\caption{NRT detections of EZOA galaxies. \label{tabdet}} \\
\hline\hline 
\noalign{\smallskip}
   \multicolumn{1}{l}{EZOA ID}           & \multicolumn{1}{l}{Class}       & \multicolumn{1}{l}{Note}             
 & \multicolumn{1}{c}{Ptg}               
 & \multicolumn{2}{c}{$v_{\rm hel}$}     & \multicolumn{2}{c}{$w_{50}$}    & \multicolumn{2}{c}{$w_{20}$}      
 & \multicolumn{2}{c}{Flux}              & \multicolumn{1}{c}{SNR}              
 & \multicolumn{1}{r}{$v_{\rm LG}$}      & \multicolumn{1}{r}{Dist}        & \multicolumn{1}{r}{$\log M_{\rm HI}$}  
 & \multicolumn{1}{l}{Final}             & \multicolumn{1}{c}{Cand}        & \multicolumn{1}{c}{$d_{\rm beam}$}
 & \multicolumn{1}{c}{RA}                & \multicolumn{1}{c}{Dec}         &  \multicolumn{1}{c}{Unc}  
 & \multicolumn{1}{c}{Gal $l$}           & \multicolumn{1}{c}{Gal $b$}     &  \multicolumn{1}{c}{$A_{\rm K}$}  
 & \multicolumn{1}{l}{Name} 
\\
   & & 
 & 
 & \multicolumn{2}{c}{[\kms ]}       & \multicolumn{2}{c}{[\kms ]}    & \multicolumn{2}{c}{[\kms ]}      
 & \multicolumn{2}{c}{[Jy\,\kms ]}   & 
 & \multicolumn{1}{r}{[\kms ]}       & \multicolumn{1}{r}{[Mpc]}      & \multicolumn{1}{r}{[\msun]}
 & & & \multicolumn{1}{c}{$b$}
 & \multicolumn{2}{c}{(J2000)}        & 
 & \multicolumn{1}{c}{[$\degr$]}     & \multicolumn{1}{c}{[$\degr$]}  &  \multicolumn{1}{c}{[mag]}  
 & 
\\
   \multicolumn{1}{l}{(1)}     & \multicolumn{1}{l}{(2)}    & \multicolumn{1}{l}{(3)}             
 & \multicolumn{1}{l}{(4)}               
 & \multicolumn{2}{c}{(5)}     & \multicolumn{2}{c}{(6)}    & \multicolumn{2}{c}{(7)}      
 & \multicolumn{2}{c}{(8)}     & \multicolumn{1}{c}{(9)}              
 & \multicolumn{1}{r}{(10)}    & \multicolumn{1}{r}{(11)}   & \multicolumn{1}{r}{(12)}  
 & \multicolumn{1}{l}{(13)}    & \multicolumn{1}{c}{(14)}   & \multicolumn{1}{c}{(15)}
 & \multicolumn{2}{c}{(16)}    &  \multicolumn{1}{c}{(17)}  
 & \multicolumn{1}{c}{(18)}    & \multicolumn{1}{c}{(19)}   &  \multicolumn{1}{c}{(20)}  
 & \multicolumn{1}{l}{(21)} 
\\
\noalign{\smallskip}
\hline
\noalign{\smallskip}
\endfirsthead
\caption{--Continued.}\\
\hline\hline
\noalign{\smallskip}
   \multicolumn{1}{l}{(1)}     & \multicolumn{1}{l}{(2)}    & \multicolumn{1}{l}{(3)}             
 & \multicolumn{1}{l}{(4)}               
 & \multicolumn{2}{c}{(5)}     & \multicolumn{2}{c}{(6)}    & \multicolumn{2}{c}{(7)}      
 & \multicolumn{2}{c}{(8)}     & \multicolumn{1}{c}{(9)}              
 & \multicolumn{1}{r}{(10)}    & \multicolumn{1}{r}{(11)}   & \multicolumn{1}{r}{(12)}  
 & \multicolumn{1}{l}{(13)}    & \multicolumn{1}{c}{(14)}   & \multicolumn{1}{c}{(15)}
 & \multicolumn{2}{c}{(16)}    &  \multicolumn{1}{c}{(17)}  
 & \multicolumn{1}{c}{(18)}    & \multicolumn{1}{c}{(19)}   &  \multicolumn{1}{c}{(20)}  
 & \multicolumn{1}{l}{(21)} 
\\
\noalign{\smallskip}
\hline
\noalign{\smallskip}
\endhead
\noalign{\smallskip}
\hline
\endfoot
J1856$-$03 & 1  & n & $A$  &  1580 $\pm$&  1 & 184 $\pm$&  2 & 200 $\pm$&  3 & 17.9 $\pm$& 0.7 & 27.4 &  1725 &  23.0 &  9.3 & 0  & p & 0   & 18 56 00.5 &$-03$ 12 21&  :  &  30.608 &$ -2.472$ & 0.40 & J1856005-031221         \\ 
J1921$+$14 & 1  & n & $A$  &  4074 $\pm$&  2 &  74 $\pm$&  4 & 99  $\pm$&  6 &  6.8 $\pm$& 0.7 & 16.3 &  4247 &  56.6 &  9.7 & 0  & d & 0   & 19 21 35.1 & +14 50 19 &  :  &  49.535 &$  0.225$ & 1.92 & J1921351+145019         \\ 
J1921$+$08 & 1  & --& $A$  &  3102 $\pm$&  3 & 125 $\pm$&  6 & 150 $\pm$&  9 &  4.9 $\pm$& 0.5 & 11.4 &  3277 &  43.7 &  9.3 & 0  & d & 0   & 19 22 00.2 & +08 18 24 &     &  43.816 &$ -2.936$ & 0.25 & J1922002+081824         \\ 
J1949$+$24 & 1  & --& $A$  &  3117 $\pm$&  2 & 248 $\pm$&  4 & 263 $\pm$&  5 &  8.0 $\pm$& 0.6 & 14.6 &  3347 &  44.6 &  9.6 & 0  & d & 0   & 19 49 53.5 & +24 11 07 &     &  60.943 &$ -1.033$ & 0.95 & J1949535+241107         \\ 
J1956$+$29 & 1  & --& $A$  &  3480 $\pm$&  3 & 180 $\pm$&  6 & 213 $\pm$&  9 &  5.8 $\pm$& 0.5 & 13.3 &  3715 &  49.5 &  9.5 & 0  & d & 0   & 19 56 24.6 & +29 35 07 &     &  66.322 &$  0.495$ & 0.79 & J1956246+293507         \\ 
J2001$+$26 & 1  & n & $A$  &  3139 $\pm$&  4 & 210 $\pm$&  9 & 226 $\pm$& 13 &  6.2 $\pm$& 0.7 & 11.9 &  3377 &  45.0 &  9.5 & 0  & d & 0   & 20 01 26.3 & +26 59 38 &  :  &  64.695 &$ -1.805$ & 0.58 & J2001263+265938         \\ 
J2029$+$31 & 1  & --& $A$  &  1045 $\pm$&  2 & 132 $\pm$&  4 & 149 $\pm$&  6 &  8.8 $\pm$& 0.7 & 14.8 &  1325 &  17.7 &  8.8 & 0  & d & 0   & 20 29 49.8 & +31 19 56 &  :  &  71.753 &$ -4.550$ & 0.25 & J2029498+311956         \\ 
J2042$+$47 & 1  & --& $A$  &  3354 $\pm$&  2 & 241 $\pm$&  5 & 253 $\pm$&  7 &  9.9 $\pm$& 0.9 & 9.5  &  3616 &  48.2 &  9.7 & 0  & d & 0   & 20 42 18.4 & +47 17 27 &     &  85.966 &$  3.112$ & 0.62 & J2042184+471727         \\
\end{longtable}
\end{landscape}
}

\twocolumn

Most outliers can be explained by the noisiness of their EBHIS profile (\eg
inclusion or exclusion of noise peaks at the edge of the profile), baseline
variations (in either profile) or possible (low-level) confusion. For
example, the latter two explanations are responsible for the high flux ratio
of 1.6 for the bright EZOA sources EZOA J2050+47 and J0437+64 ($F_{\rm NRT}
\sim 16$\,\Jykms), respectively.  In one case (EZOA J0314+64), the EBHIS detection
was found at the edge of the cube, losing part of the signal of its
\HI\ disc ($\Delta v = -5$\,\kms , $\Delta w_{50} = -41$\,\kms , $F_{\rm
  NRT} / F_{\rm EBHIS} =  1.6$. In another case (EZOA J0253+55), the
\HI\ disc is resolved in the NRT observation ($\Delta v = 2$\,\kms ,
$\Delta w_{50} = -7$\,\kms , $F_{\rm NRT} / F_{\rm EBHIS} = 0.5$. Where no
cross-match is found (green circles), the galaxy may not be situated very
close to the centre of the NRT beam, resulting in a lower detected
flux. While this obviously influences the flux ratios (the mean improves to
$0.92 \pm 0.03$ when green circles are excluded), it does not cause any
significant deviations in velocity and line widths.

\section{Discussion}   \label{discussion}  

\subsection{Recovery rate}   \label{recov}  

Our NRT target list contains both weaker and potential EZOA detections. 
The deeper NRT follow-up observations allow now a better assessment regarding the
various EZOA classifications. This will be very useful for future source
detection efforts with EBHIS and other, similar, surveys.

\begin{table}
\caption[]{NRT recovery rate of EZOA sources.}
\label{tabrecov}
\begin{tabular}{lrrrr}
\hline
\noalign{\smallskip}
EZOA class    &    1  &   3+   &    3   &   $3-$ \\
\noalign{\smallskip}
\hline
\noalign{\smallskip}
EZOA sources total &  111  &   40   &   19   &   $-$  \\
NRT observed sources      &   36  &   25   &   16   &   22   \\
NRT detected sources      &   36  &   22   &   10   &    4   \\
Inconclusive/possible$^a$ &   $-$ &   $1$  &   $1$  &   $1$  \\
NRT recovery rate  &  100\%&   88\% &   63\% &   18\% \\ 
\noalign{\smallskip}
\hline
\noalign{\smallskip}
Total confirmed sources  &  -- &   37   &   13   &   -- \\
Total confirmation rate         &  -- &   93\% &   68\% &   -- \\ 
\noalign{\smallskip}
\hline
\multicolumn{5}{l}{\small{$^a$: considered as not detected when calculating
rates}}
\end{tabular}
\end{table}

Table~\ref{tabrecov} gives the NRT recovery rate as a function of the EZOA
source classification, which was based on the SNR, appearance in the
spectral cube, form of the profile, and other information, like the
proximity of RFI or baseline variations. Most of the detections selected
for this follow-up study have a low SNR, hence an intrinsically lower
reliability. The recovery rates validate our internal classification scheme
and show a clear offset between, on the one hand, solid and marginal EZOA
detections (classes 1, 3+ and 3, in order of declining confidence levels)
and, on the other, the lowest class $3-$, which indicates possible
detections (not included in the published EZOA catalogue). The striking
decrease from a recovery of 63\% for class 3 to 18\% for $3-$ confirms the
classification scheme, including the label `possible' for the lowest class
detections.

These recovery rates, however, should not be considered as final. The NRT
target list only covered a subsample of all EZOA sources.  The total
recovery rate for the class~1 EZOA detections should be stable at
100\%\footnote{Given the full recovery of all lower SNR detections, we can
  assume that all other class~1 detections to be equally reliable.}, the
rates of class 3+ and 3 sources increase if we include those we did not
observe. Of the 40 class 3+ EZOA sources, 15 were not observed since they
all had previously been detected in \HI . Including these raises the total
recovery rate to 93\%. Similarly for class~3: if we include the three
sources we did not observe, the total recovery rate increases to 68\%.

Further investigating the recovery rates, we checked the three sources with
a class 3+ classification in EZOA that were not detected with the NRT: the
two non-detections were indeed borderline cases to a class~3. The third
EZOA source, a firmer class 3+ detection, is inconclusive since three 
NRT pointings were affected by GPS RFI and one good observation had a high rms.

Regarding the class $3-$ sources, it is interesting to note that all four
confirmed sources had been independently identified in EBHIS data by a
neural network software package (\citealt{floer15}; see also the discussion
in \citealt{schroeder19}). The package uses an automated source finding
algorithm to detect candidate sources, then uses a neural network to
classify these as detections or non-detections. The classification
procedure had been improved by using the EZOA sources found by visual
inspection of the \HI\ cubes as well as a list of rejects from an older
version of the neural network catalogue \citep{schroeder19}. This resulted
in 15 new sources that had not been found visually, but only three of them
were subsequently judged to be sufficiently strong and acceptable to be
included in the class~$3-$ list (the majority were too faint for making a
decision during the visual inspection). \citet{schroeder19} give a detailed
discussion on the completeness and reliability of this automated source
catalogue, mentioning a 100\% reliability of sources with SNR\,$> 8$,
while the four confirmed sources have SNRs in the range 5 to
7. In contrast, only one of the possible sources that had been found by
eye\footnote{Incidentally, it was also found by the improved neural
  network} was recovered by the NRT observations, indicating a higher
reliability of machine learning classifications at the low SNR end.

It is also worth noting that secondary information, like the presence or
absence of a (credible) cross-match\footnote{Taking into account extinction
  levels and star densities in the field in combination with the \HI\ mass
  and line width of the detection, that is, do we expect to see this
  source?}  is useful in estimating the reliability of an \HI\ detection,
with the caveat that at low SNRs the flux, and with it the \HI\ mass, have a higher
uncertainty. To investigate this further, we separate the recovery rates
into sources with and without a cross-match in
Table~\ref{tabrecov2}. The differences are striking for class~3 and
3$-$ sources. The numbers need to be taken with a grain of salt, however,
because (a) we did not take into account whether a cross-match is
expected to be visible or not (given the foreground extinction and star
density at its location\footnote{To note, \citet{staveley16} give a discussion on the
  dependence of finding a cross-match with Galactic longitude, that is,
  closer to the bulge it becomes more difficult to find a cross-match due
  to higher extinctions and star densities.}), and (b) we are
dealing with small number statistics (less than 10 in many cases). However,
an \HI\ detection with a cross-match
is more likely to be reliable for all classes.

\begin{table}
\caption[]{Recovery rate of EZOA sources with and without a cross-match.}
\label{tabrecov2}
\begin{tabular}{lrrrr}
\hline
\noalign{\smallskip}
EZOA class    &    1  &   3+   &    3   &   $3-$ \\
\noalign{\smallskip}
\hline
\noalign{\smallskip}
Total                &  100\%&   88\% &   63\% &   18\% \\ 
With cross-match     &  100\%&   94\% &   88\% &   43\% \\ 
Without cross-match  &  100\%&   71\% &   43\% &    7\% \\ 
\noalign{\smallskip}
\hline
\end{tabular}
\end{table}

\begin{figure*}  
\centering
\includegraphics[width=0.9\textwidth]{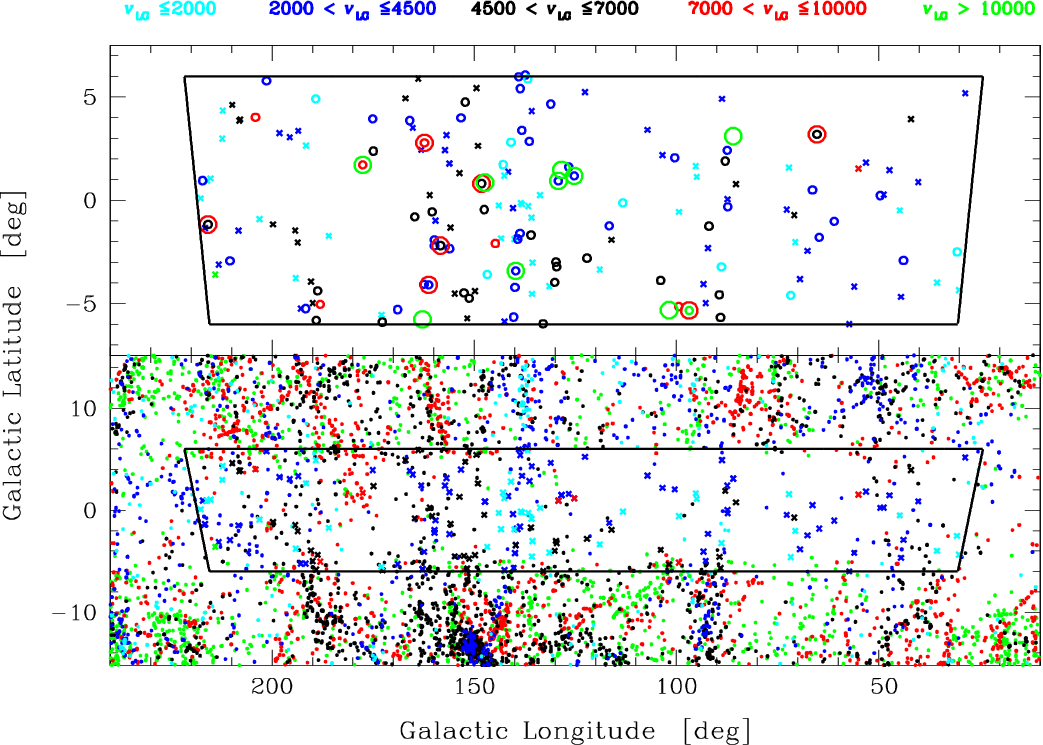}
\caption{On-sky distribution of \HI\ detections in Galactic coordinates;
  the symbols are colour-coded as a function of radial velocity (see labels
  at the top). Top panel: The original 170 EZOA catalogue galaxies (circles
  depict those re-observed with the NRT); NRT non-detections are
  encircled in red, the newly found NRT detections in green. Bottom panel:
  EZOA sources based on the updated list (crosses), combined with galaxies
  listed in HyperLEDA (dots). For a better assessment of the large-scale
  structures, we have increased the range in Galactic latitude.
}
\label{figmap}
\end{figure*}

\subsection{New detections }   \label{newdet} 

Due to the higher sensitivity of the NRT observations and the large
north-south extent of the NRT beam, we have found five serendipitous
detections. They are listed at the end of Table~\ref{tabdet}. For three of
them we could identify a cross-match. Only one of these (EZOA J0253+55B) has
a previously known redshift \citep{kraan18}.

Adding these to the detections of the four formerly class~$3-$ sources, we
have nine detections in total that were not listed in the original EZOA
catalogue. On the other hand, seven EZOA sources could not be confirmed and
need to be removed from the catalogue. The top panel of Fig.~\ref{figmap}
shows the on-sky distribution of all 170 EZOA sources (colour-coded by
velocity; circles depict NRT observations) with the new NRT detections and
non-detections indicated by green and red circles, respectively. The bottom
panel shows the distribution of the updated EZOA catalogue sources together
with velocity data available in the literature (through HyperLEDA). The
EZOA detections clearly trace several large scale structures across the gap
of the ZoA; see \citet{schroeder19} for a detailed discussion of these
structures.

\subsection{RFI mitigation test results} \label{rfi} 

Our RFI mitigation procedure relies on the main assumption that time-line
series of voltages recorded in the NRT receiver chain from non-variable
cosmic sources have normal, that is, Gaussian distributions, whereas any
artificial (\eg RFI) signal cannot have a Gaussian distribution, since it
carries coherent information. The corresponding time-line series of power
spectral values in each frequency channel have a $\chi^2$ distribution with
two degrees of freedom. Thus, averaged power
spectral values from $N$ individual power spectra are $\chi^2$ distributed
with $2N$ degrees of freedom. By virtue of the Central Limit Theorem, for
$N\gtrsim25$ the $\chi^2$ distribution approaches a normal
distribution. Therefore, we configured the WIBAR spectrometer in such a way
that it produces time-line series of averaged power spectral values which
fulfil the $N\gtrsim25$ condition.

We regard RFI signals as statistical
outliers, that is points lying outside the bulk of the normal distribution
in each frequency channel. To detect RFI, we compare non-robust estimators
of location ($EoL$), such as the mean, and of scale ({\it EoS}), such as the
standard deviation, to robust {\it EoL} and {\it EoS} which are immune to outliers.
For the latter, we respectively chose the median as {\it EoL}, and compared
results obtained with Median Absolute Deviation (MAD) and its alternatives
{\it Sn} and {\it Qn} (see \citealt{rousseeuw93}), as
{\it EoS}. To remove RFI, we applied a $5 \times 3 Sn$ recursive clipping
process in each frequency channel, since we found {\it Sn} to provide the best
compromise between efficiency and computing time for large samples (see
\citealt{belleval19} for further details).

Results of our RFI mitigation test observations are shown here only for the
two galaxies where the test results add pertinent new information to the standard
NRT ACRT and the EZOA data; the observations of the other test sources were
marred by instrumental problems  (cf.\ Sect.~\ref{observations}).

\begin{figure}[h]  
\centering
\includegraphics[width=0.49\textwidth]{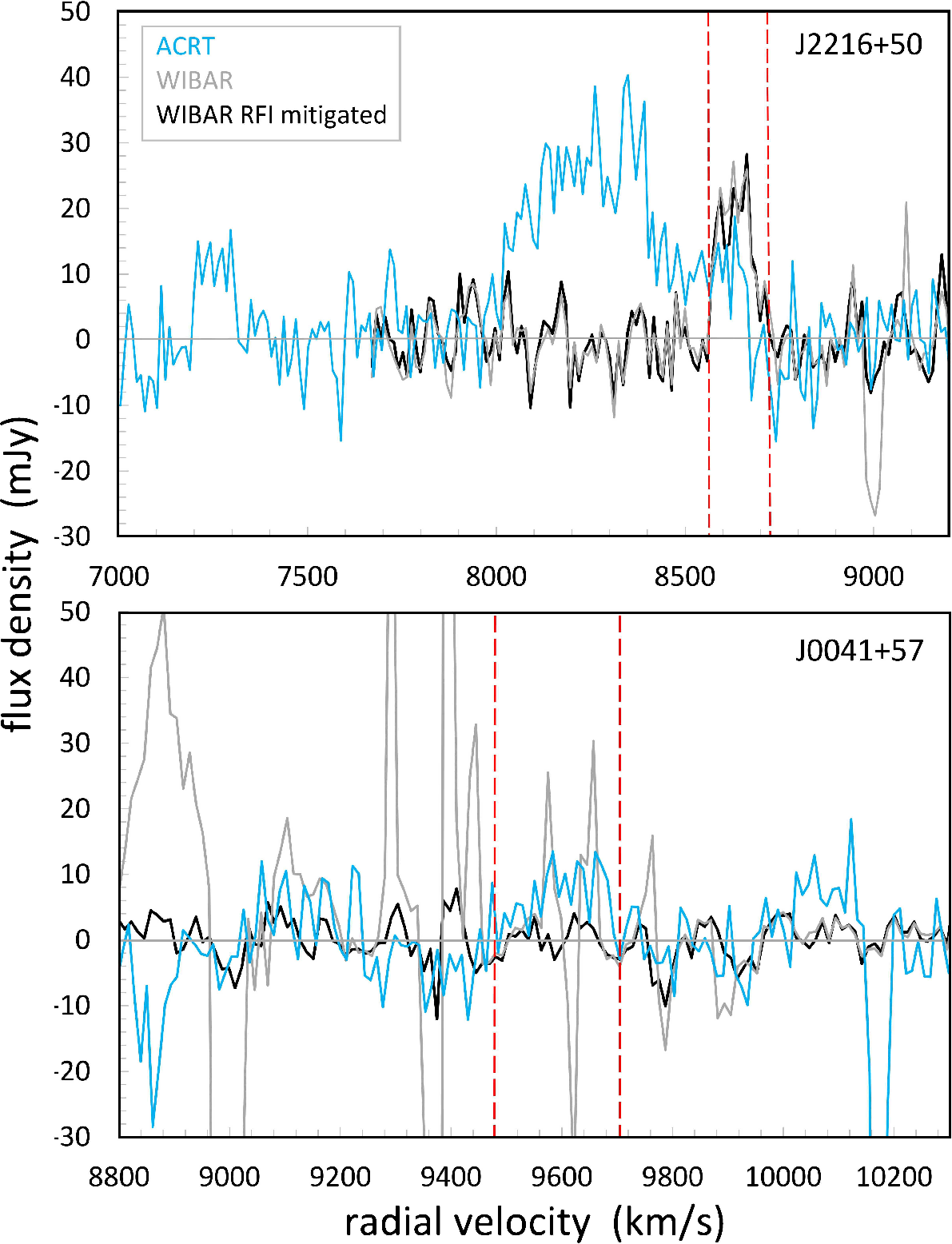}
\caption{Results of RFI mitigation test observations of two objects. Shown
  are standard \nan\ ACRT auto-correlator \HI\ line spectra (in blue) and
  spectra obtained with the WIBAR broadband receiver, both without RFI
  mitigation (in grey) and after RFI mitigation using the ROBEL software package 
  (in black). Shown is flux density (in mJy) as a function of heliocentric
  radial velocity in the optical framework, {\it cz} (in \kms). The velocity
  resolution is 11.1\,\kms\ for the WIBAR spectra, and 10.3\,\kms\ for the
  standard ACRT spectra. The pairs of red dotted vertical lines indicate
  the profile widths of the EBHIS spectra.
}
\label{figRFItests}
\end{figure}

Figure~\ref{figRFItests} shows the results for the two sources, EZOA
J2216+50 and J0047+57, which are discussed below. Spectra obtained with the
WIBAR receiver before and after processing with the ROBEL RFI
mitigation package are shown as grey and black lines, respectively. Also
shown (in blue) are the standard ACRT auto-correlator spectra.

The WIBAR spectrum of EZOA J2216+50 after RFI mitigation (black lines)
shows that the ACRT spectrum is dominated by residuals of broad GPS L3 RFI
around 8200\,\kms. These unwanted emissions could be successfully mitigated
in the WIBAR data to reveal a clear detection within the EZOA profile
velocity range (red dotted lines). A comparison of the \HI\ line parameters
with the EBHIS parameters is given in Table~\ref{tabrficomp}. While the
agreement in velocity and line width is very good, the flux is lower than
expected since we observed here the EBHIS position (pointing $B$) instead
of the actual cross-match at $1.0b$ distance.

\begin{table}
\caption[]{Comparison of line parameters in the RFI mitigation test for
  EZOA J2216+50. }
\label{tabrficomp}
\begin{tabular}{lccccc}
\hline
\noalign{\smallskip}
Observation & $v_{\rm hel}$ & $w_{50}$ & Flux & rms & SNR \\
            & \kms          & \kms     & Jy\,\kms & mJy & \\
\noalign{\smallskip}
\hline
\noalign{\smallskip}
ROBEL & 8628 & 107 & 2.4 & 3.6 & 6.4 \\ 
EBHIS & 8626 & 107 & 6.2 & 9.0 & 7.1 \\
\noalign{\smallskip}
\hline
\end{tabular}
\end{table}

For EZOA J0041+57, only narrow RFI signals are detected in the WIBAR
spectrum (grey lines) within the EZOA profile velocity range, which could
be mitigated (black line), resulting in an estimated rms noise level of 3.0 mJy
(compared to 11.1 mJy in the EBHIS spectrum). This indicates that the
reported EZOA detection was due to RFI signals, which is not unexpected,
considering that EZOA J0047+57 was only a possible (class $3-$) detection.

\section{Summary} \label{summary}  

In this paper, we present follow-up 21\,cm \HI\ line observations made with
the 100\,m-class NRT of selected sources in the northern ZoA from the
EZOA catalogue \citep{schroeder19}. These sources were
extracted form the shallow, first pass, version of the EBHIS blind
\HI\ survey performed with the Effelsberg 100\,m radio telescope
(\citealt{kerp11} and \citealt{winkel10}). Our selection criteria include low
SNR ($\lessapprox 10$), no or ambiguous optical or NIR counterpart, and
noisy profiles, resulting in 77 EZOA sources. In addition, we included
22 candidate detections that were deemed too unreliable to be published in
the EZOA paper.

While the rms noise level of the shallow survey EBHIS cubes is
23\,mJy\,beam$^{-1}$, the median rms level of the EZOA detections is
10.4\,mJy\,beam$^{-1}$ \citep{schroeder19}, and the NRT follow-up
observations have an almost three times lower median rms level of
3.6\,mJy\,beam$^{-1}$. Of the 99 observed sources, we confirmed 72,
while three yielded inconclusive results. Most of the non-detections (17
out of 24) were potential EZOA sources with a low reliability. In
addition, due to the higher sensitivity of our NRT observations as well as
the large north-south extent of the NRT beam, we have found five
serendipitous detections.

We investigated how the EZOA classification scheme of the detections
(based, among others, on appearance in the spectral cube and the profile)
correlates with the recovery rate. We find that the classification scheme
is very effective with a 100\% recovery rate for definite detections, 93\%
and 68\% for two sets of marginal detections, and only 18\% for detections
considered to have a low likelihood to be real. We also found that for potential
detections, the presence or absence of a suitable cross-match candidate is
a useful indicator for reliability.

Finally, we have updated the EZOA catalogue parameters for the 75 detected
EZOA sources, removed the non detections and added the four confirmed EZOA
candidate detections. The five serendipitous detections are listed separately.

Though only two sources yielded useful results, the RFI mitigation test
observations proved useful. Unwanted emission due to RFI was successfully
mitigated in both spectra, showing in one case a solid detection, while in
the other case any previously noted emission could be attributed to RFI. In
both cases the rms noise was significantly reduced.

\begin{acknowledgements}

We want to thank Matt Lehnert for his help with the analysis of
spectra. This research has made use of the Pan-STARRS1 Surveys public science
archive and the HyperLeda database\footnote{http://leda.univ-lyon1.fr}. The
\nan{} Radio Observatory is operated by the Paris Observatory, associated
with the French Centre National de la Recherche Scientifique. RKK wishes to
thank the South African NRF for their financial support.

\end{acknowledgements}



\bibliographystyle{aa} 
\bibliography{ZoA_bibfile} 

\begin{thebibliography}{53}
\expandafter\ifx\csname natexlab\endcsname\relax\def\natexlab#1{#1}\fi

\bibitem[{{Belleval}(2019)}]{belleval19}
{Belleval}, C. 2019, {Robust statistics applied to radio astronomy: RFI
  mitigation and automated spectral line detection for broadband surveys, PhD
  Thesis,} (Paris Observatory)

\bibitem[{{Chamaraux} {et~al.}(1990){Chamaraux}, {Cayatte}, {Balkowski}, \&
  {Fontanelli}}]{chamaraux90}
{Chamaraux}, P., {Cayatte}, V., {Balkowski}, C., \& {Fontanelli}, P. 1990,
  A\&A, 229, 340

\bibitem[{{Courtois} {et~al.}(2009){Courtois}, {Tully}, {Fisher}, {Bonhomme},
  {Zavodny}, \& {Barnes}}]{courtois09}
{Courtois}, H.~M., {Tully}, R.~B., {Fisher}, J.~R., {et~al.} 2009, \aj, 138,
  1938

\bibitem[{{Donley} {et~al.}(2005){Donley}, {Staveley-Smith}, {Kraan-Korteweg},
  {Islas-Islas}, {Schr{\"o}der}, {et~al.}}]{donley05}
{Donley}, J.~L., {Staveley-Smith}, L., {Kraan-Korteweg}, R.~C., {et~al.} 2005,
  AJ, 129, 220

\bibitem[{{Dressler} {et~al.}(1987){Dressler}, {Lynden-Bell}, {Burstein},
  {Davies}, {Faber}, {Terlevich}, \& {Wegner}}]{dressler87}
{Dressler}, A., {Lynden-Bell}, D., {Burstein}, D., {et~al.} 1987, ApJ, 313, 42

\bibitem[{{Fl{\"o}er}(2015)}]{floer15}
{Fl{\"o}er}, L. 2015, {Automated source extraction for the next generation of
  neutral hydrogen surveys, PhD Thesis,} (University of Bonn)

\bibitem[{{Fl{\"o}er} {et~al.}(2014){Fl{\"o}er}, {Winkel}, \& {Kerp}}]{floer14}
{Fl{\"o}er}, L., {Winkel}, B., \& {Kerp}, J. 2014, \aap, 569, A101

\bibitem[{{Focardi} {et~al.}(1984){Focardi}, {Marano}, \&
  {Vettolani}}]{focardi84}
{Focardi}, P., {Marano}, B., \& {Vettolani}, G. 1984, \aap, 136, 178

\bibitem[{{Goedhart} {et~al.}(2024){Goedhart}, {Cotton}, {Camilo}, {Thompson},
  {Umana}, {Bietenholz}, , {et~al.}}]{goedhart24}
{Goedhart}, S., {Cotton}, W.~D., {Camilo}, F., {et~al.} 2024, \mnras, 531, 649

\bibitem[{{Haynes} {et~al.}(2018){Haynes}, {Giovanelli}, {et~al.}}]{alfalfa}
{Haynes}, M.~P., {Giovanelli}, R., {et~al.} 2018, \apj, 861, 49

\bibitem[{{Henning} {et~al.}(1998){Henning}, {Kraan-Korteweg}, {Rivers},
  {Loan}, {Lahav}, \& {Burton}}]{henning98}
{Henning}, P.~A., {Kraan-Korteweg}, R.~C., {Rivers}, A.~J., {et~al.} 1998, AJ,
  115, 584

\bibitem[{{Henning} {et~al.}(2008){Henning}, {Springob}, {Day}, {Minchin},
  {Momjian}, {Catinella}, {Muller}, {Koribalski}, {Masters}, {Pantoja},
  {Putman}, {Rosenberg}, {Schneider}, \& {Staveley-Smith}}]{henning08}
{Henning}, P.~A., {Springob}, C.~M., {Day}, F., {et~al.} 2008, in American
  Institute of Physics Conference Series, Vol. 1035, The Evolution of Galaxies
  Through the Neutral Hydrogen Window, ed. R.~{Minchin} \& E.~{Momjian},
  246--248

\bibitem[{{Huchra} {et~al.}(2012){Huchra}, {Macri}, {Masters}, {Jarrett},
  {Berlind}, \& {Calkins}}]{huchra12}
{Huchra}, J.~P., {Macri}, L.~M., {Masters}, K.~L., {et~al.} 2012, ApJS, 199, 26

\bibitem[{{Hudson} {et~al.}(2004){Hudson}, {Smith}, {Lucey}, \&
  {Branchini}}]{hudson04}
{Hudson}, M.~J., {Smith}, R.~J., {Lucey}, J.~R., \& {Branchini}, E. 2004,
  MNRAS, 352, 61

\bibitem[{{Kerp} {et~al.}(2011){Kerp}, {Winkel}, {Ben Bekhti}, {Fl{\"o}er}, \&
  {Kalberla}}]{kerp11}
{Kerp}, J., {Winkel}, B., {Ben Bekhti}, N., {Fl{\"o}er}, L., \& {Kalberla},
  P.~M.~W. 2011, Astronomische Nachrichten, 332, 637

\bibitem[{{Koribalski} {et~al.}(2004){Koribalski}, {Staveley-Smith}, {Kilborn},
  {Ryder}, {Kraan-Korteweg}, {Ryan-Weber}, {Ekers}, {Jerjen},
  {et~al.}}]{koribalski04}
{Koribalski}, B.~S., {Staveley-Smith}, L., {Kilborn}, V.~A., {et~al.} 2004, AJ,
  128, 16

\bibitem[{{Kraan-Korteweg} {et~al.}(2017){Kraan-Korteweg}, {Cluver}, {Bilicki},
  {Jarrett}, {Colless}, {Elagali}, {B{\"o}hringer}, \& {Chon}}]{kraan17}
{Kraan-Korteweg}, R.~C., {Cluver}, M.~E., {Bilicki}, M., {et~al.} 2017, \mnras,
  466, L29

\bibitem[{{Kraan-Korteweg} {et~al.}(2002){Kraan-Korteweg}, {Henning}, \&
  {Schr{\"o}der}}]{kraan02}
{Kraan-Korteweg}, R.~C., {Henning}, P.~A., \& {Schr{\"o}der}, A.~C. 2002, \aap,
  391, 887

\bibitem[{{Kraan-Korteweg} {et~al.}(2018){Kraan-Korteweg}, {van Driel},
  {Schr{\"o}der}, {Ramatsoku}, \& {Henning}}]{kraan18}
{Kraan-Korteweg}, R.~C., {van Driel}, W., {Schr{\"o}der}, A.~C., {Ramatsoku},
  M., \& {Henning}, P.~A. 2018, \mnras, 481, 1262

\bibitem[{{Kraan-Korteweg} {et~al.}(1996){Kraan-Korteweg}, {Woudt}, {Cayatte},
  {Fairall}, {Balkowski}, \& {Henning}}]{kraan96}
{Kraan-Korteweg}, R.~C., {Woudt}, P.~A., {Cayatte}, V., {et~al.} 1996, \nat,
  379, 519

\bibitem[{{Kurapati} {et~al.}(2024){Kurapati}, {Kraan-Korteweg}, {Pisano},
  {Chen}, {Rajohnson}, {Steyn}, {Frank}, {Serra}, {Goedhart}, \&
  {Camilo}}]{kurapati24}
{Kurapati}, S., {Kraan-Korteweg}, R.~C., {Pisano}, D.~J., {et~al.} 2024,
  \mnras, 528, 542

\bibitem[{{Lang} {et~al.}(2003){Lang}, {Boyce}, {Kilborn}, {Minchin}, {Disney},
  {Jordan}, {Grossi}, {Garcia}, {Freeman}, {Phillipps}, \& {Wright}}]{lang03}
{Lang}, R.~H., {Boyce}, P.~J., {Kilborn}, V.~A., {et~al.} 2003, VizieR Online
  Data Catalog, 734, 20738

\bibitem[{{Louw} {et~al.}(2024){Louw}, {Kurapati}, {Pisano}, \&
  {Kraan-Korteweg}}]{louw24}
{Louw}, A., {Kurapati}, S., {Pisano}, D.~J., \& {Kraan-Korteweg}, R. 2024, in
  IAU General Assembly, 1043

\bibitem[{{Macri} {et~al.}(2019){Macri}, {Kraan-Korteweg}, {Lambert}, {Alonso},
  {Berlind}, {Calkins}, {Erdo{\u{g}}du}, {Falco}, {Jarrett}, \&
  {Mink}}]{macri19}
{Macri}, L.~M., {Kraan-Korteweg}, R.~C., {Lambert}, T., {et~al.} 2019, \apjs,
  245, 6

\bibitem[{{Masters} {et~al.}(2014){Masters}, {Crook}, {Hong}, {Jarrett},
  {Koribalski}, {Macri}, {Springob}, \& {Staveley-Smith}}]{masters14}
{Masters}, K.~L., {Crook}, A., {Hong}, T., {et~al.} 2014, \mnras, 443, 1044

\bibitem[{{Matthews} \& {van Driel}(2000)}]{matthews00}
{Matthews}, L.~D. \& {van Driel}, W. 2000, A\&AS, 143, 421

\bibitem[{{McIntyre} {et~al.}(2015){McIntyre}, {Henning}, {Minchin}, {Momjian},
  \& {Butcher}}]{mcintyre15}
{McIntyre}, T.~P., {Henning}, P.~A., {Minchin}, R.~F., {Momjian}, E., \&
  {Butcher}, Z. 2015, \aj, 150, 28

\bibitem[{{Meyer} {et~al.}(2004){Meyer}, {Zwaan}, {Webster}, {Staveley-Smith},
  {Ryan-Weber}, {Drinkwater}, {Barnes}, {Howlett}, {et~al.}}]{meyer04}
{Meyer}, M.~J., {Zwaan}, M.~A., {Webster}, R.~L., {et~al.} 2004, MNRAS, 350,
  1195

\bibitem[{{Monnier Ragaigne} {et~al.}(2003){Monnier Ragaigne}, {van Driel},
  {Schneider}, {Balkowski}, \& {Jarrett}}]{monnier03c}
{Monnier Ragaigne}, D., {van Driel}, W., {Schneider}, S.~E., {Balkowski}, C.,
  \& {Jarrett}, T.~H. 2003, A\&A, 408, 465

\bibitem[{{O'Neil}(2004)}]{oneil04b}
{O'Neil}, K. 2004, AJ, 128, 2080

\bibitem[{{Paturel} {et~al.}(2003){Paturel}, {Theureau}, {Bottinelli},
  {Gouguenheim}, {Coudreau-Durand}, {Hallet}, \& {Petit}}]{paturel03}
{Paturel}, G., {Theureau}, G., {Bottinelli}, L., {et~al.} 2003, A\&A, 412, 57

\bibitem[{{Planck Collaboration} {et~al.}(2016){Planck Collaboration},
  {Aghanim}, {Ashdown}, {Aumont}, {Baccigalupi}, {Ballardini}, {Band ay},
  {Barreiro}, {Bartolo}, {et~al.}}]{planck16b}
{Planck Collaboration}, {Aghanim}, N., {Ashdown}, M., {et~al.} 2016, \aap, 596,
  A109

\bibitem[{{Rajohnson} {et~al.}(2024){Rajohnson}, {Kraan-Korteweg}, {Chen},
  {Frank}, {Steyn}, {Kurapati}, {Pisano}, {Staveley-Smith}, {Serra},
  {Goedhart}, \& {Camilo}}]{rajohnson24}
{Rajohnson}, S. H.~A., {Kraan-Korteweg}, R.~C., {Chen}, H., {et~al.} 2024,
  \mnras, 531, 3486

\bibitem[{{Ramatsoku} {et~al.}(2016){Ramatsoku}, {Verheijen}, {Kraan-Korteweg},
  {J{\'o}zsa}, {Schr{\"o}der}, {Jarrett}, {Elson}, {van Driel}, {de Blok}, \&
  {Henning}}]{ramatsoku16}
{Ramatsoku}, M., {Verheijen}, M.~A.~W., {Kraan-Korteweg}, R.~C., {et~al.} 2016,
  \mnras, 460, 923

\bibitem[{{Rivers} {et~al.}(1999){Rivers}, {Henning}, \&
  {Kraan-Korteweg}}]{rivers99}
{Rivers}, A.~J., {Henning}, P.~A., \& {Kraan-Korteweg}, R.~C. 1999, \pasa, 16,
  48

\bibitem[{{Rousseeuw} \& {Croux}(1993)}]{rousseeuw93}
{Rousseeuw}, P.~J. \& {Croux}, C. 1993, Journal of the American Statistical
  Association, 88, 1273

\bibitem[{{Said} {et~al.}(2016){Said}, {Kraan-Korteweg}, {Staveley-Smith},
  {Williams}, {Jarrett}, \& {Springob}}]{said16a}
{Said}, K., {Kraan-Korteweg}, R.~C., {Staveley-Smith}, L., {et~al.} 2016,
  \mnras, 457, 2366

\bibitem[{{Sanchez-Barrantes} {et~al.}(2019){Sanchez-Barrantes}, {Henning},
  {McIntyre}, {Momjian}, {Minchin}, {Rosenberg}, {Schneider}, {Staveley-Smith},
  {van Driel}, {Ramatsoku}, {Butcher}, \& {Vaez}}]{sanchez19}
{Sanchez-Barrantes}, M., {Henning}, P.~A., {McIntyre}, T., {et~al.} 2019, \aj,
  158, 234

\bibitem[{{Schr{\"o}der} {et~al.}(2019{\natexlab{a}}){Schr{\"o}der},
  {Fl{\"o}er}, {Winkel}, \& {Kerp}}]{schroeder19}
{Schr{\"o}der}, A.~C., {Fl{\"o}er}, L., {Winkel}, B., \& {Kerp}, J.
  2019{\natexlab{a}}, \mnras, 489, 2907

\bibitem[{{Schr{\"o}der} {et~al.}(2009){Schr{\"o}der}, {Kraan-Korteweg}, \&
  {Henning}}]{schroeder09}
{Schr{\"o}der}, A.~C., {Kraan-Korteweg}, R.~C., \& {Henning}, P.~A. 2009, \aap,
  505, 1049

\bibitem[{{Schr{\"o}der} {et~al.}(2019{\natexlab{b}}){Schr{\"o}der}, {van
  Driel}, \& {Kraan-Korteweg}}]{schroeder19b}
{Schr{\"o}der}, A.~C., {van Driel}, W., \& {Kraan-Korteweg}, R.~C.
  2019{\natexlab{b}}, \mnras, 482, 5167

\bibitem[{{Schr{\"o}der} {et~al.}(2021){Schr{\"o}der}, {van Driel}, \&
  {Kraan-Korteweg}}]{schroeder21}
{Schr{\"o}der}, A.~C., {van Driel}, W., \& {Kraan-Korteweg}, R.~C. 2021,
  \mnras, 503, 5351

\bibitem[{{Springob} {et~al.}(2005){Springob}, {Haynes}, {Giovanelli}, \&
  {Kent}}]{springob05}
{Springob}, C.~M., {Haynes}, M.~P., {Giovanelli}, R., \& {Kent}, B.~R. 2005,
  ApJS, 160, 149

\bibitem[{{Springob} {et~al.}(2016){Springob}, {Hong}, {Staveley-Smith},
  {Masters}, {Macri}, {Koribalski}, {Jones}, {Jarrett}, {Magoulas}, \&
  {Erdo{\u{g}}du}}]{springob16}
{Springob}, C.~M., {Hong}, T., {Staveley-Smith}, L., {et~al.} 2016, \mnras,
  456, 1886

\bibitem[{{Staveley-Smith} {et~al.}(2016){Staveley-Smith}, {Kraan-Korteweg},
  {Schr{\"o}der}, {Henning}, {Koribalski}, {Stewart}, \& {Heald}}]{staveley16}
{Staveley-Smith}, L., {Kraan-Korteweg}, R.~C., {Schr{\"o}der}, A.~C., {et~al.}
  2016, \aj, 151, 52

\bibitem[{{Steyn} {et~al.}(2024){Steyn}, {Kraan-Korteweg}, {Rajohnson},
  {Kurapati}, {Chen}, {Frank}, {Serra}, {Staveley-Smith}, {Camilo}, \&
  {Goedhart}}]{steyn24a}
{Steyn}, N., {Kraan-Korteweg}, R.~C., {Rajohnson}, S. H.~A., {et~al.} 2024,
  \mnras, 529, L88

\bibitem[{{Tully} {et~al.}(2019){Tully}, {Pomar{\`e}de}, {Graziani},
  {Courtois}, {Hoffman}, \& {Shaya}}]{tully19}
{Tully}, R.~B., {Pomar{\`e}de}, D., {Graziani}, R., {et~al.} 2019, \apj, 880,
  24

\bibitem[{{Tully} {et~al.}(2008)}]{tully08}
{Tully}, R.~B. {et~al.} 2008, ApJ, 676, 184

\bibitem[{{van Driel} {et~al.}(2016){van Driel}, {Butcher}, {Schneider},
  {Lehnert}, {Minchin}, {Blyth}, {Chemin}, {Hallet}, {Joseph}, {Kotze},
  {Kraan-Korteweg}, {Olofsson}, \& {Ramatsoku}}]{vandriel16}
{van Driel}, W., {Butcher}, Z., {Schneider}, S., {et~al.} 2016, \aap, 595, A118

\bibitem[{{van Driel} {et~al.}(2009){van Driel}, {Schneider}, {Kraan-Korteweg},
  \& {Monnier Ragaigne}}]{vandriel09}
{van Driel}, W., {Schneider}, S.~E., {Kraan-Korteweg}, R.~C., \& {Monnier
  Ragaigne}, D. 2009, A\&A, 505, 29

\bibitem[{{Winkel} {et~al.}(2010){Winkel}, {Kalberla}, {Kerp}, \&
  {Fl{\"o}er}}]{winkel10}
{Winkel}, B., {Kalberla}, P.~M.~W., {Kerp}, J., \& {Fl{\"o}er}, L. 2010, \apjs,
  188, 488

\bibitem[{{Wong} {et~al.}(2006){Wong}, {Ryan-Weber}, {Garcia-Appadoo},
  {Webster}, {Staveley-Smith}, {Zwaan}, {Meyer}, {Barnes}, {et~al.}}]{wong06}
{Wong}, O.~I., {Ryan-Weber}, E.~V., {Garcia-Appadoo}, D.~A., {et~al.} 2006,
  MNRAS, 371, 1855

\bibitem[{{Woudt} {et~al.}(2004){Woudt}, {Kraan-Korteweg}, {Cayatte},
  {Balkowski}, \& {Felenbok}}]{woudt04}
{Woudt}, P.~A., {Kraan-Korteweg}, R.~C., {Cayatte}, V., {Balkowski}, C., \&
  {Felenbok}, P. 2004, \aap, 415, 9

\end{thebibliography}


\begin{appendix}

\section{Notes on individual galaxies}  \label{appnotes}  

In the following we discuss individual cases where the \HI\ line emission
shows a complex profile or its origin is ambiguous. When discussing possible
contamination of a profile by a source close to the target, we
often prefer to express angular distances in units of the NRT beam radius
$b$ to better understand relative parameters like flux ratios.

To help with identifying potential cross-matches, we have compiled
Table~\ref{tabgal} with uncatalogued galaxies around those EZOA detections
that we have observed at the NRT (and that are large enough for be relevant
for our discussion). These were found while examining multi-wavelength images 
in the search for cross-matches (see \citealt{schroeder19}). We
usually only list galaxies within the EZOA position uncertainty radius of
$3\farcm5$, though in cases of suspected in-beam confusion also galaxies
further away are listed. Furthermore, we list previously catalogued
galaxies where we improved the position, since these are important for our
discussion. We note that all previously published galaxies near EZOA
detections that are not listed in this table can be found in the CDS Simbad
and NED databases. The columns are (1) EZOA ID (without the prefix); (2)
galaxy ID (previously uncatalogued galaxies are given J2000.0-coordinate
IDs); (3) equatorial (J2000.0) coordinates, with an uncertainty flag in
Col.\ 4 if the centre of the galaxy could not easily be determined; (5)
angular distance from the nominal EZOA position in arcminutes (for
comparison, the EZOA position uncertainty is $3\farcm5$); and (6) a comment
where necessary.

\noindent {\bf J1856$-$03:} There is a possible galaxy faintly visible in
deep NIR images at (RA,Dec) $=(18^h56^m00\fs5,-03{\degr}12'21'')$ at a
distance of $2\farcm2$ from the EZOA position. Its large and diffuse
appearance matches the profile well. The NRT observation of this position
shows an improved profile with even peaks and a high flux ratio of 1.2,
which seems to confirm this candidate.

\noindent {\bf J1921$+$14:} The possible cross-match candidate has been confirmed.

\noindent {\bf J2001$+$26:} There are three galaxies in the field, one of
which (2MASX J20010969+2655338) was observed by \citet{kraan18} with the \nrt\
but not detected. We observed now the other two and can confirm the one closer
to the EZOA position (pointing~$A$) as the cross-match. 

\noindent {\bf J2102$+$46:} The possible cross-match candidate has been confirmed.

\noindent {\bf J2125$+$48:} There is no cross-match for this marginal (SNR
$=5.2$) EZOA detection. We observed it with four pointings. We have
possible detections at RA $= 21^h25^m36\fs9$ (pointing~$A$) and RA $=
21^h25^m16\fs0$ (pointing~$D$), albeit with less than the expected flux and
showing only the lower velocity horn. However, the baseline of all
observations is affected by GPS residuals, the rms is high (between 4 and
7) and the SNR is low (around 3). Ideally we need more observations to
improve the SNRs, but for the time being we assume that the detection is
only a possible one and might not be real.

\noindent {\bf J2131$+$43:}
We observed the most likely cross-match (pointing~$A$) and obtained a
clearly confused profile, about twice as wide as the EBHIS profile. To the
south, at a distance of $3\farcm1$ from the EZOA position (and $0.6b$ from
our pointing~$A$), is a tight group of galaxies (2MASX J21312321+4336182)
that was observed by \citet{paturel03}, listed in our table as pointing
$B$\footnote{It was also observed by \citet{masters14}, but we prefer a
direct comparison of two NRT detections because of the NRT's unusual beam
form.}. The flux at pointing~$A$ is marginally higher than at pointing~$B$,
and also pointing~$A$'s peak flux is about 10\,mJy higher. We conclude that
the most of the EZOA detection comes from our cross-match at (RA,Dec) $=
(21^h31^m17\fs9,+43{\degr}39'03'')$. 2MASX J21312321+4336182 likely
contributes, and its low-velocity part is, in fact, visible in the EBHIS
spectrum, though it is strongly affected by baseline variations.

\noindent {\bf J2216$+$50:}
We observed both the cross-match position (pointing~$A$) and the EZOA
position (pointing~$B$). Both are affected by RFI. We could confirm the
cross-match, but the measured flux is unreliable due to insufficient RFI
subtraction. Pointing~$B$ was also observed with the WIBAR receiver to test
the RFI mitigation software, resulting in a clearer profile
(Sect..~\ref{rfi}). However, the pointing is $1.0b$ from the cross-match
position, resulting in a flux ratio of 0.4 as compared to the uncertain
flux ratio of 0.7 found for pointing~$A$.

\noindent {\bf J2230$+$51:}
This possible EZOA detection was not detected in the pointings towards two
possible cross-match candidates. However, we have a serendipitous detection
at $v=5806$\,\kms , which is like due to the targeted cross-match candidate
(pointing~$A$).

\noindent {\bf J2237$+$53:}
The cross-match is a galaxy pair (2MASX J22370662+5357049 is nearly edge
on, while 2MASX J22370933+5358339 is fairly inclined), both of which
contribute to the \HI\ detection. We observed the former, while the latter
(at $0.25b$) was observed previously both with the NRT \citep{paturel03}
and the Green Bank 300-foot telescope (GBT; \citealt{courtois09}); we have
added them to our table for comparison. Our NRT observation shows a clear
peak at $v<5100$\,\kms , which is not obvious in the EBHIS profile.
Comparing the two NRT profiles we can tell by the relative peak heights
that the profile at $5100 < v < 5400$\,\kms\ comes from J22370662+5357049,
while J22370933+5358339 covers $5200 < v < 5550$\,\kms . Thus the EZOA
detection comes mainly from J22370933+5358339. Note that \citet{paturel03}
only measure the profile from $v>5200$\,\kms\ (same as the EBHIS profile),
while \citet{courtois09} give parameters for the full profile.

\noindent {\bf J2329$+$61:}
There is no obvious cross-match for this detection. We observed the EZOA
position (pointing~$A$) as well as positions about $1b$
($\sim\!\!1\farcm8$) to the west (pointing~$B$) and east (pointing~$C$),
respectively, to cover the EBHIS positional uncertainty area.  Only
pointing $A$ at the EZOA position shows a marginal detection, confirming
the EZOA detection, but with a much lower flux (note, though, that the
EBHIS profile has a low SNR of 4.8). The cross-match is likely found north
or south of the EZOA position at a larger distance.

\noindent {\bf J0041$+$57:}
The NRT spectrum looks like a {\it bona fide} detection at $v=9547$\,\kms ,
but it lies within a velocity range where RFI occurs frequently.
Simultaneous observation with the WIBAR receiver and subsequent RFI
mitigation shows that the apparent profile is entirely due to RFI (see
Sect.~\ref{rfi}).

\noindent {\bf J0112$+$63:} 
We observed this marginal EZOA detection, which that has no cross-match,
with three pointings, covering the EBHIS positional uncertainty area. The
profile is marginal at pointing~$A$ and stronger at pointing~$B$ though the
flux is still lower than the EBHIS value ($\sim\!50$\%). The profile shape
appears less lopsided than in the EBHIS detection, which agrees with the
cross-match lying closer to position~$B$. We decided to re-investigate NIR
images of this area and found that at (RA,Dec)
$=(01^h12^m12\fs0,+63{\degr}57'14'')$ the DSS2-$I$ and 2MASS-$K$ band
images show a diffuse patch which could be a low surface brightness galaxy
(this agrees with the high extinction of $A_K=0\fm6$). Pan-STARRS1 images
confirm this to be a galaxy. It is $0.8b$ west of pointing~$B$, which
agrees with the measured flux ratio of 0.5.

Pointing~$A$ also shows a serendipitous detection at $v=7503$\,\kms , for
which no cross-match is visible either.

\noindent {\bf J0125$+$64:} There is no cross-match for this detection. 
Our detection at the EZOA position has about half the EBHIS flux ($\sim
50$\%). We conclude that the cross-match lies either east or west near the
edge of the NRT beam.

\noindent {\bf J0141$+$63:}
Without a possible cross-match we observed the EZOA position for this
possible EBHIS detection. We detect a marginal profile, albeit with a
$\sim\!30$\% lower flux. We also tried to observe a galaxy visible with
{\it WISE} (pointing~$B$) but the observation failed. Due to the low NRT
flux (though both detections have a low SNR), we expect that the
cross-match is likely nearer the edges of the NRT and EBHIS beams (more
likely in the east-west direction).

\noindent {\bf J0147$+$63:} There is no obvious cross-match for this
detection. We observed the EZOA position (pointing~$A$) as well as a
position at RA $= 01^h47^m33^s$ about $1.4b$ to the east (pointing
$B$). Further southeast, $5\farcm1$ (or $2.4b$) away, lies 2MASX
J01474890+6305128, which was previously observed and detected by us
(see \citealt{kraan18}) and is listed as pointing~$C$ in Table~\ref{tabobs}
for convenience. All three observations show a detection, albeit with
varying profile shapes. We conclude that the EBHIS detection is likely a
confused profile with an invisible galaxy close to the EZOA position as the
main contributor and 2MASX J01474890+6305128 as the confusing partner
(contributing mainly at the high velocity end of the profile).

We also found a serendipitous detection at $v=7515$\,\kms\ in pointing~$A$,
but no candidate for it.

\noindent {\bf J0213$+$66:}
\citet{schroeder19} reported two galaxies in the field without being able to
tell which could be the cross-match. We observed both and find that the
galaxy at (RA,Dec) $=(02^h13^m11\fs5,+66{\degr}12'28'')$ (pointing~$B$) is
the cross-match. We have also detected the other galaxy at (RA,Dec)
$=(02^h13^m33\fs4,+66{\degr}10'39'')$ as a weak, low-velocity shoulder at
$v\, \approxgt\ 4020$\,\kms , which is not visible in the EBHIS
spectrum. This indicates that the two galaxies could form a pair.

\noindent {\bf J0233$+$58:}
We confirm the cross-match with 2MASX J02333595+5836438 but also report
emission at the high-velocity end of the profile, $v\, \approxlt\
4720$\,\kms , which is not visible in the EBHIS spectrum. It may come from
a companion galaxy, 2MASX J02334017+5836388.

\noindent {\bf J0252$+$62:}
As the cross-match 2MASX J02520392+6235505 lies at a large distance from
the EZOA position ($4\farcm0$ north), it is likely that another galaxy
contributes to the profile. We observed both the cross-match and a galaxy
about $0.9b$ to the South of it (and $4\farcm4$ south of the EZOA
position), visible on DSS and {\it WISE} images. For comparison, we also
observed the EZOA position (pointing~$C$). We find that at both galaxy
positions the flux is slightly higher than at the EZOA position. We
conclude that both cross-matches contribute to the EBHIS detection.

\noindent {\bf J0253$+$55:}
The NRT/EBHIS flux ratio at the probable cross-match candidate position
(pointing~$A$) is low at 0.5, though it should be noted that the EBHIS
profile is noisy with an rms of 14\,mJy. When we observed the EZOA position
(pointing~$B$), we found an even lower flux ratio of 0.3. This indicates
that the cross-match is either the observed candidate or an invisible
galaxy close to it.

Furthermore, a galaxy pair at $\sim \!4\farcm4$ from the EZOA position had
been observed previously by us with the NRT (\citealt{kraan18}): 2MASX
J02531475+5528143 and 2MASX J02531969+5529140 are also listed in the table
as pointing~$C$ and $D$, respectively. The pair was detected at
$v=4462$\,\kms\ and $4337$\,\kms , respectively, with fluxes too low for
detection in the EBHIS survey (though the former is also visible in our
spectrum of pointing~$A$). However, both observations also show the EZOA
detection at $v=3809$\,\kms\ albeit at a lower flux. \citet{kraan18} also
observed a third, uncatalogued galaxy further north (pointing~$E$).  The
separations between the EZOA cross-match candidate and these three galaxies
are $1.0b$, $1.4b$ and $1.1b$, respectively, and all three show only the
higher velocity peak at $v\sim 3820$\,\kms , while our observation of the
candidate shows a narrow double horn with a second (slightly higher) peak
at $v\sim 3700$\,\kms , which, in turn, is the main peak in pointing $B$.

We conclude that the cross-match has a large (resolved) \HI\ disc, whose
lower-velocity end lies close to the EZOA position while the high-velocity
end points towards the galaxy pair. This would also explain the overall
lower fluxes measured at the NRT.

\noindent {\bf J0308$+$64:}
This EZOA detection is of a local dwarf galaxy (at a distance of 9.8
Mpc; \citealt{schroeder19}) that has no stellar counterpart.  We observed
the EZOA position and detected a strong signal at a slightly lower flux
(70\%). The fact that the NRT profile shows two peaks of equal height
indicates that the galaxy must lie well within the telescope beam. We
subsequently searched Pan-STARRS1 images and found a diffuse galaxy only
$0\farcm25$ from the EZOA position, which, indeed, is also visible (though
not recognisable as a galaxy) on DSS2-R, at (RA,Dec)
$=(03^h08^m27\fs3,+64{\degr}59'37'')$.

\noindent {\bf J0314$+$64:}
Since the EZOA detection is located at the edge of the EBHIS cube, its
profile is noisy and its position and \HI\ parameters are uncertain. Our
observation of the cross-match 2MASX J03151297+6452348 shows a profile
which is $\sim\!80$\,\kms\ wider than could be estimated from the noisy
EBHIS profile.

\noindent {\bf J0324$+$60:} 
The search for a cross-match revealed a diffuse patch on the DSS2-R
image that could be a late-type dwarf galaxy but was too uncertain to be
listed as cross-match in \citet{schroeder19}. We observed this position
with a flux ratio of 0.7. Though the flux is slightly lower than expected,
we can confirm the cross-match since we also noted an elongated, diffuse
LSB galaxy on Pan-STARRS1 images. At a distance of 18.3\,Mpc we can assume
that the \HI\ disc is resolved by the NRT beam, which explains the lower
flux.

\noindent {\bf J0328$+$53} and {\bf J0328+53B:}
EZOA J0328$+$53 ($v=8538$\,\kms ) is a noisy EBHIS detection with a double
horned profile, and EZOA J0328+53B ($v=11\,952$\,\kms ) is a possible
single-peak detection. The two detections are separated by $\sim1\farcm9$.
There is one possible cross-match candidate visible to the south (at $1.1b$
and $0.5b$ from the two EZOA source positions, respectively), which we
observed (both pointings $A$) along with the EZOA J0328$+$53 position
(pointings $B$). Pointing~$A$ is affected by RFI in the velocity range
$8500-9500$\,\kms\ and is therefore inconclusive regarding the EZOA
J0328$+$53 detection, while EZOA J0328+53B is clearly not detected.
Pointing~$B$ shows baseline variations in the range $8000-9000$\,\kms\ and
no recognisable \HI\ profile, nor does it show a detection for EZOA
J0328+53B.

\noindent {\bf J0332$+$58:}
\citet{schroeder19} mention a possible galaxy at
(RA,Dec) $=(03^h32^m48\fs4,+58{\degr}14'55'')$, which could be the
cross-match. Our observation of this position (pointing~$A$) shows a
detection, but with too low a flux (60\%). We also observed the EZOA
position (pointing~$B$), which shows a much stronger profile. The NRT/EBHIS
flux ratio at pointing~$B$ is 0.9, indicating that the actual cross-match
is very close to the $B$ position. We thus give a new positional
uncertainty for the EZOA detection with $0\fm2$ in RA and $1\farcm4$ in Dec
(referring to the inner ten percent of the NRT beam).

\noindent {\bf J0349$+$53:}
This marginal, single-peak EZOA detection is confirmed (pointing~$A$) with
an NRT/EBHIS flux ratio of 0.5, which implies that the cross-match lies
near the edge of the NRT beam in the east-west direction.

\noindent {\bf J0358$+$47:}
The NRT profile at the position of the cross-match ZOAG G152.52-04.48 shows
a sloped high-velocity shoulder (at $v \approxlt\ 5700$\,\kms ). To check
if this could be due to confusion, we also observed a galaxy at $0.4b$.
This profile, though, shows both a lower shoulder and a lower total flux,
and we therefore exclude confusion.

\noindent {\bf J0431$+$45:}
Though this was considered a firm, although weak and noisy, detection
by \citet{schroeder19}, we could not confirm it with three pointings
covering the EBHIS positional uncertainty area.

\noindent {\bf J0433$+$39:}
We observed the cross-match of this possible EZOA source, $3\farcm0$ from
the EZOA position. We detect a strong double horn profile with equal peaks
but with a lower flux (60\%). The very noisy EBHIS profile is about twice
as wide as the NRT profile but seems to include a noise peak, which would
explain the lower flux in our NRT detection.

\noindent {\bf J0437$+$43:}
\citet{schroeder19} give a galaxy pair as cross-match. We observed the
northern component (2MASX J04370228+4356359; pointing~$A$) but were not
able to observe the eastern one (2MASX J04370607+4355349, a more inclined,
earlier type spiral). The two galaxies are only $0.4b$ apart. The NRT
profile looks clearly confused and wider than the EBHIS profile, which is
quite noisy and, basically, shows only the two highest peaks. The galaxy
was also observed by \citet{springob05} with the GBT. Their profile is
heavily smoothed and shows few features, but it is obvious that is shows
emission at higher velocities than our NRT profile (at $v\, \approxlt\
4100$\,\kms ). Since the east-west GBT HPBW is larger than the NRT's by a
factor of about three, we assume that they have detected more flux from the
companion galaxy, whose disc is oriented in an east-west direction, and
that the NRT beam therefore would not see all of it. We conclude that both
galaxies are detected, with the emission at the higher velocity end of the
profile coming from 2MASX J04370607+4355349.

\noindent {\bf J0437$+$54:}
We observed a pair of galaxies: 2MASX J04373506+5414339 (pointing~$A$) is
fairly face-on, at $\vopt=5369$\,\kms\ \citep{huchra12} and is considered
the probable EZOA cross-match, while 2MASX J04374087+5415389 (pointing~$B$)
is almost edge-on, at $\vopt=5653$\,\kms\ \citep{huchra12}. Our NRT
observations show an overall stronger profile for the former, while the
profile of pointing~$B$ is weaker and seems to have a high-velocity
shoulder. We conclude that both galaxies are detected but only 2MASX
J04373506+5414339 contributes to the noisy EBHIS profile.

\noindent {\bf J0438$+$44:}
\citet{schroeder19} noted the close-by galaxy UGC\,3108 at $d$\,=\,$4\farcm4$
from the EZOA position but excluded its contribution to the signal. We
observed the adopted cross-match 2MASX J04381402+4407596 ($1.5b$ from
UGC\,3108) and confirm the EZOA detection. UGC\,3108 was observed
by \citet{springob05} with the GBT, where it shows a wider profile starting
at $v<3800$\,\kms\ (listed in our table as pointing~$B$). The GBT profile
shows our detection as a pronounced high velocity horn at $v = 4000 -
4100$\,\kms . Although we cannot exclude that the EBHIS and NRT
observations both see part of the \HI\ disc of UGC\,3108, the contribution
from our cross-match is clearly dominating.

\noindent {\bf J0446$+$44:}
We observed the EZOA position since there is no cross-match. The detected
signal shows an NRT/EZOA flux ratio of 1.1 and the profile shows symmetric
peaks. We conclude that the cross-match is well within the NRT beam, that
is, very close to the EZOA position. We thus give a new positional
uncertainty for the EZOA detection with $0\fm2$ in RA and $1\farcm3$ in Dec
(referring to the inner ten percent of the NRT beam).

\noindent {\bf J0455$+$34:} The possible cross-match candidate has been confirmed.

\noindent {\bf J0457$+$39:}
The NRT observation at the position of this possible EZOA detection shows a
$3\sigma$ detection with only 10\% of the EBHIS flux recovered. The two
pointings east and west of the EZOA position show no detection at all. It
is possible that the cross-match lies further north or south (note, though,
that the north-south NRT HPBW of 22$'$ is twice that of EBHIS), or the
detection may be spurious. More observations are needed to confirm this
weak signal.

\noindent {\bf J0546$+$31:}
Apart from the cross-match 2MASX J05462224+3155264, there are several
smaller, uncatalogued, galaxies visible on the images, possibly forming a
galaxy group. We observed the three most prominent ones and can confirm
that 2MASX J05462224+3155264 (pointing~$A$) is indeed the cross-match. We
also detect at least one of the smaller ones (pointing~$C$, with a
prominent peak at $v=7600$\,\kms ), and cannot exclude low level
contamination of the NRT detections by one or more of the others. The EBHIS
parameters are likely unaffected due to the lower sensitivity.

\noindent {\bf J0548$+$19:}
The cross-match is a close galaxy pair, 2MASX J05485392+1911467. We
observed the main component (pointing~$A$, an inclined large spiral), which
shows a clearly confused profile, possibly due to the smaller companion, an
earlier type spiral at $d=0\farcm25$. The EBHIS detection does not show
confusion, but it is very weak, and the sloped low-velocity shoulder
($v \approxlt\ 5700$\,\kms ) seen with the NRT is likely lost in the
noise. There is also a large galaxy far to the south-east (2MASX
J05490625+1904314; at $1.6b$), which we observed (pointing~$B$) to exclude
the possibility that the elongated NRT beam might have picked up emission
from it. Although this spectrum shows a detection, it is too weak to
account for the strong low-velocity shoulder visible in pointing
$A$. Instead, this is likely an off-beam detection of our galaxy pair.

\noindent {\bf J0636$+$00:}
Deep NIR images reveal the cross-match to be a galaxy triplet: pointing~$A$
targets 2MASX J06362668+0055433, which consists of an inclined spiral
galaxy and a peculiar-looking companion (fragmented, or possibly a tight
group of small -- or distant -- galaxies), while  pointing~$B$ 
($0\farcm8$ to the west) 2MASX J06362361+0055513 appears more face-on. The
observed \HI\ fluxes are about the same. Since the two pointings are
separated by $0.4b$, we would expect a notable difference between the two
fluxes if only one source were detected. We therefore conclude that both
(or all three) galaxies are detected by both the NRT and EBHIS.

\begin{figure}
\centering
\includegraphics[width=0.49\textwidth]{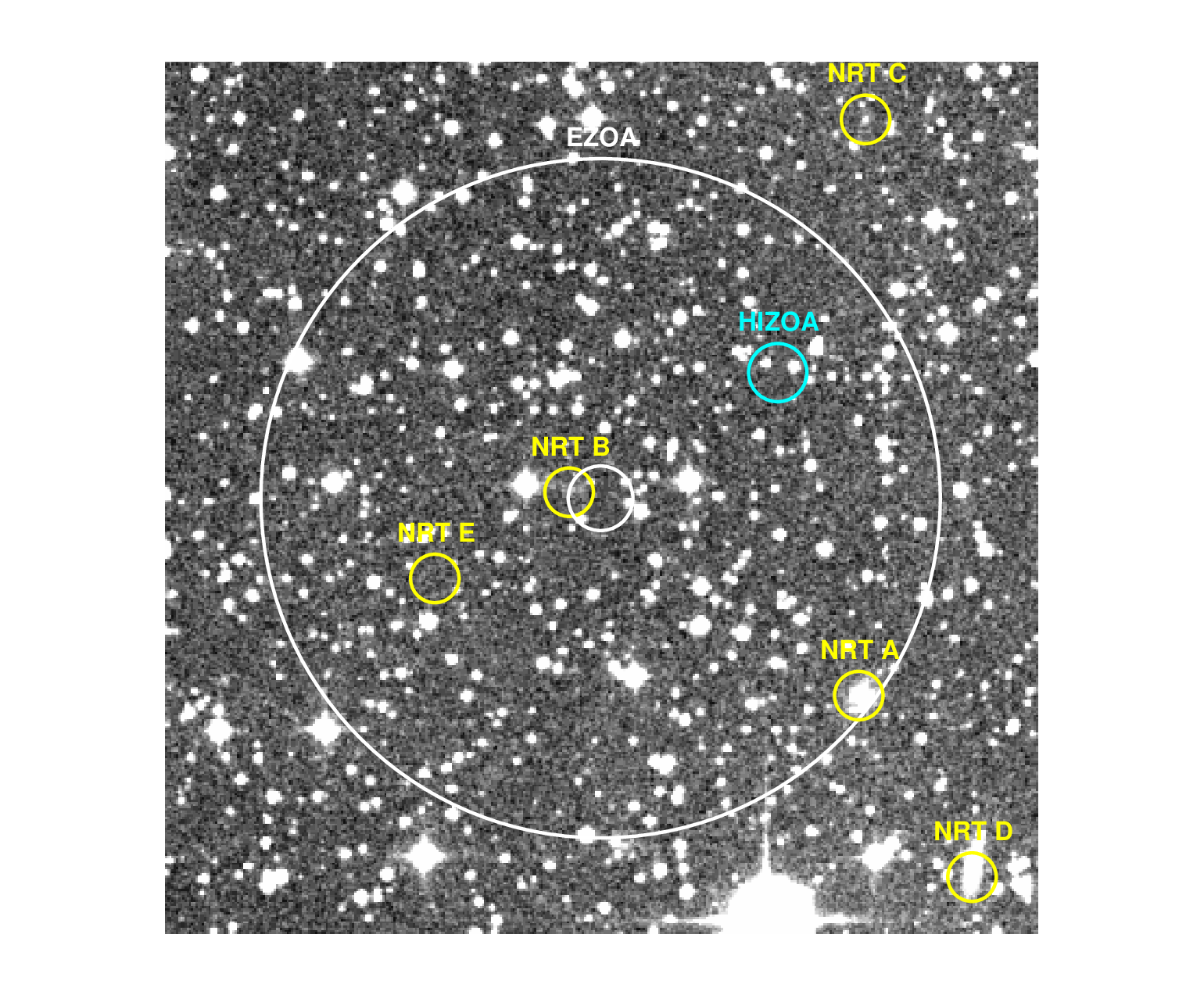}
\caption{$R$-band $10\arcmin \times 10\arcmin$ image of EZOA
J0649$+$09. Labelled are the EZOA detection and positional uncertainty
circle (white), the HIZOA detection (cyan) and the five NRT pointings
(yellow).
}
\label{e200_10}
\end{figure}

\noindent {\bf J0649$+$09:}
The most likely cross-match is 2MASX J06493148+0939437 at a rather large
distance of $d$\,=\,$3\farcm3$ from the EZOA position. We observed this
galaxy (pointing~$A$), but the detection is dominated by a strong
OFF-position detection (from about $20'$ east). Near that position lies
UGC\,3565, which was detected by \citet{springob05} with the Arecibo
telescope in the velocity range
$\sim7250-7600$\,\kms . This implies that half of our profile is affected
by it, and only the NRT peak at $v\sim7620$\,\kms\ lies just outside the
profile of UGC\,3565. We also observed four other galaxies in the area (see
Fig.~\ref{e200_10}).

The peak at $v\sim7620$\,\kms\ is fainter for pointing~$C$ and appears of
similar strength in pointings $A$ and $D$ (2MASX J06492673+0937525) though
both are likely diminished (in different proportions) by the OFF-position
profile. On the other hand, the two easterly pointings, $B$ and $E$, are
unaffected by the OFF-detection. They show detections in the range
$v=7300-7600$\,\kms\ where the profile at $v=7500-7600$\,\kms\ is slightly
stronger in pointing~$E$ and the emission at $v=7300-7500$\,\kms\ is likely
due to the target of pointing~$B$.

We conclude that the galaxies at pointing~$B$ and $E$ are both detected in
the velocity range $v=7300 - 7600$\,\kms . Based on the spectra, we cannot
distinguish whether one or both of the 2MASX galaxies has emission at
$v\approxlt\ 7620$\,\kms . However, 2MASX J06493148+0939437 at pointing~$A$
is more likely to be a strong \HI\ detection since it is closer to both the
EZOA position and the nearby HIZOA detection \citep{donley05}. The latter
shows a likely confused profile with a peak at $v\sim 7550$\,\kms , which
we do not see in any of the NRT profiles due to the OFF-position detection.
The galaxy at pointing~$C$ does not seem to have any emission in the
velocity range. As to the EBHIS detection, it does not appear confused but
the profile is noisy and its SNR of 5.4 is low. The EBHIS profile starts at
$v=7350$\,\kms , which means the galaxy at pointing~$B$ contributes only
little, and most of the emission seems to come from 2MASX
J06493148+0939437.

\noindent {\bf J0652$-$03:}
This marginal EZOA detection could not be confirmed. The ATNF HIZOA data
archive\footnote{https://www.atnf.csiro.au/research/multibeam/release}
shows a similar peak, which contributed to its inclusion into the EZOA
catalogue. However, none of the three NRT pointings show any indication of
a detection (the peak flux density should be around 100\,mJy): pointing~$A$ is the
most likely cross-match, pointing~$B$ is the EZOA position, and pointing
$C$ was set at a search position between these two to form a continuous
coverage of this area with the NRT beams.

\noindent {\bf J0702$-$03:}
This detection was also found by \citet{staveley16} as HIZOA J0702$-$03B at
(RA,Dec) $=(07^h02^m34\fs5,-03{\degr}18'30'')$, about $1\farcm9$ north of
the EZOA position. There are two small cross-match candidates about $0.6b$
to the West. We observed the northern, more likely one (pointing~$A$) as
well as the EZOA position (pointing~$B$). The flux ratio at pointing~$A$ is
0.9, as opposed to 0.6 at the EZOA position. We can thus confirm that the
cross-match lies close to pointing~$A$, though both candidates appear too
small for a galaxy at a radial distance of $\sim 32$\,Mpc and an \HI\ mass
of $\log(M_{\rm{HI}} / M_\odot) $ = 9.1. To note, there is a bright star close by, which could
prevent us from seeing a possible LSB disk or from finding the cross-match.


\onecolumn
{\scriptsize
\begin{longtable}{
llrr@{\extracolsep{0.mm}}c@{\extracolsep{0.mm}}rl
}
\caption{Uncatalogued galaxies around EZOA sources and previously catalogued galaxies with improved positions.
\label{tabgal}} \\
\hline\hline 
\noalign{\smallskip}
EZOA ID & galaxy ID & \multicolumn{1}{c@{\extracolsep{0.mm}}}{RA}           & \multicolumn{1}{c@{\extracolsep{0.mm}}}{Dec} &
unc     & dist  & Comment \\
 &  & \multicolumn{2}{c@{\extracolsep{0.mm}}}{(J2000) } & flag & $[']$ & \\
   \multicolumn{1}{c@{\extracolsep{0.mm}}}{(1)}    & \multicolumn{1}{c@{\extracolsep{0.mm}}}{(2)}         
 & \multicolumn{2}{c@{\extracolsep{0.mm}}}{(3)}    & \multicolumn{1}{l@{\extracolsep{0.mm}}}{(4)}           
 & \multicolumn{1}{l@{\extracolsep{0.mm}}}{(5)}    & \multicolumn{1}{l@{\extracolsep{0.mm}}}{(6)}   \\
\noalign{\smallskip}
\hline
\noalign{\smallskip}
\endfirsthead
\caption{--Continued.}\\
\hline\hline
\noalign{\smallskip}
EZOA ID & galaxy ID & \multicolumn{1}{c@{\extracolsep{0.mm}}}{RA}           & \multicolumn{1}{c@{\extracolsep{0.mm}}}{Dec} &
unc     & dist  & Comment \\
 &  & \multicolumn{2}{c@{\extracolsep{0.mm}}}{(J2000) } & flag & $[']$ & \\
   \multicolumn{1}{c@{\extracolsep{0.mm}}}{(1)}    & \multicolumn{1}{c@{\extracolsep{0.mm}}}{(2)}         
 & \multicolumn{2}{c@{\extracolsep{0.mm}}}{(3)}    & \multicolumn{1}{l@{\extracolsep{0.mm}}}{(4)}           
 & \multicolumn{1}{l@{\extracolsep{0.mm}}}{(5)}    & \multicolumn{1}{l@{\extracolsep{0.mm}}}{(6)}   \\
\noalign{\smallskip}
\hline
\noalign{\smallskip}
\endhead
\noalign{\smallskip}
\hline
\endfoot
J1856$-$03&       J1856005$-$031221 &  18 56 00.5 &$-03$ 12 21 &  : &  2.2  & Possible LSB galaxy                          \\ 
J1921+08  &       J1922002+081824   &  19 22 00.2 &  +08 18 24 &    &  1.6  &                                              \\ 
          &       J1921433+082034   &  19 21 43.3 &  +08 20 34 &    &  3.1  &                                              \\ 
J1921+14  &       J1921351+145019   &  19 21 35.1 &  +14 50 19 &  : &  2.9  &                                              \\ 
J1949+24  &       J1949535+241107   &  19 49 53.5 &  +24 11 07 &    &  0.6  &                                              \\ 
J1956+29  &       J1956246+293507   &  19 56 24.6 &  +29 35 07 &    &  1.0  &                                              \\ 
J2001+26  &       J2001263+265938   &  20 01 26.3 &  +26 59 38 &  : &  1.8  &                                              \\ 
          &       J2001267+265536   &  20 01 26.7 &  +26 55 36 &    &  3.6  &                                              \\
\end{longtable}
}

\twocolumn

\section{Observations table and \HI\ spectra } \label{apponline} 

\onecolumn
{\scriptsize
\begin{landscape}
\begin{longtable}{
 l@{\extracolsep{1.mm}}                                     l@{\extracolsep{1.mm}}                                   
 c@{\extracolsep{0.mm}}                                     r@{\extracolsep{1.5mm}}                                  r@{\extracolsep{2.mm}}  
 r@{\extracolsep{0.5mm}}                                    c@{\extracolsep{0.5mm}}                                  r@{\extracolsep{3.mm}}  
 r@{\extracolsep{0.5mm}}                                    c@{\extracolsep{0.5mm}}                                  l@{\extracolsep{2.mm}}  
 r@{\extracolsep{0.5mm}}                                    c@{\extracolsep{0.5mm}}                                  l@{\extracolsep{0.mm}}  
 c@{\extracolsep{1.mm}}                                     
 c@{\extracolsep{1.mm}}                                     r@{\extracolsep{2.mm}}                                   r@{\extracolsep{1.mm}}  
 c@{\extracolsep{1.mm}}                                     l@{\extracolsep{0.mm}}                                   l@{\extracolsep{0.mm}}  
 c@{\extracolsep{1.mm}}                                     c@{\extracolsep{1.mm}}                                   c@{\extracolsep{0.mm}}  
 r@{\extracolsep{4.mm}}                                     
 r@{\extracolsep{0.5mm}}                                    c@{\extracolsep{0.mm}}                                   r@{\extracolsep{2.mm}}  
 r@{\extracolsep{0.5mm}}                                    c@{\extracolsep{-2.mm}}                                  r@{\extracolsep{2.mm}}  
 r@{\extracolsep{0.5mm}}                                    c@{\extracolsep{0.mm}}                                   r@{\extracolsep{2.mm}}  
 r@{\extracolsep{0.5mm}}                                    c@{\extracolsep{0.5mm}}                                  l@{\extracolsep{0.mm}}  
 r@{\extracolsep{1.mm}}                                     
}
\caption{\HI\ observations of EZOA sources. \label{tabobs}} \\
\end{longtable}
\end{landscape}
}

\twocolumn

\begin{figure*}
\centering
\includegraphics[width=0.98\textwidth]{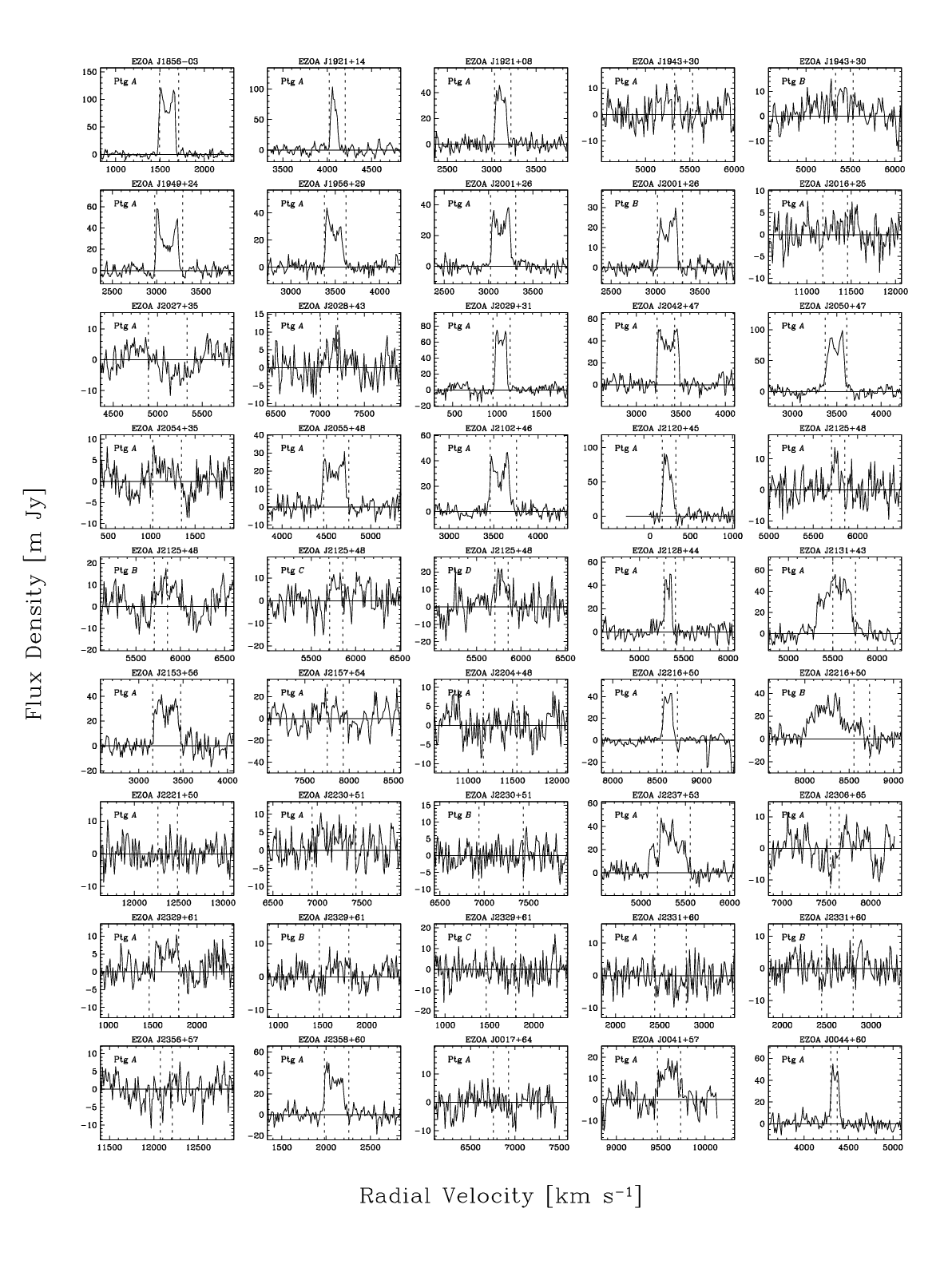}
\vspace{-1cm}
\caption{NRT 21cm \HI\ line spectra of EZOA source observations. The
  velocity resolution is 10\,\kms. The pointing ID (see Table~\ref{tabobs}) is 
  indicated for each spectrum. To help the comparison with the EBHIS
  detections, the dashed vertical lines in each spectrum indicate the
  velocity range of the EBHIS detection, which was excluded in the EBHIS baseline
  fit. 
}
\label{figspectra}
\end{figure*}

\newpage
\clearpage 

\begin{figure*}
\centering
\addtocounter{figure}{-1}
\includegraphics[width=0.98\textwidth]{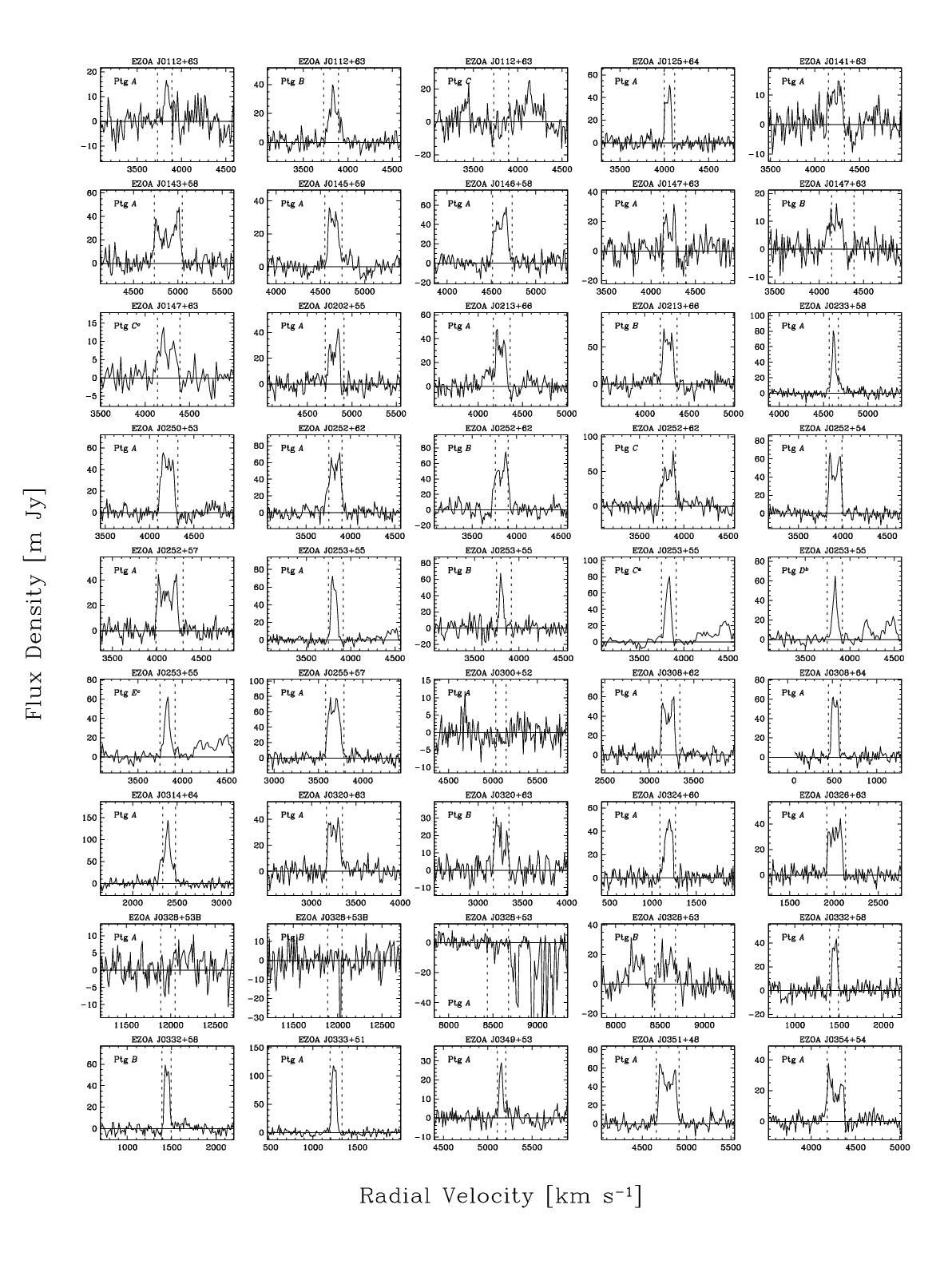}
\vspace{-1cm}
\caption{NRT 21cm \HI\ line spectra -- {continued.}
}
\end{figure*}

\newpage
\clearpage 

\begin{figure*}
\centering
\addtocounter{figure}{-1}
\includegraphics[width=0.98\textwidth]{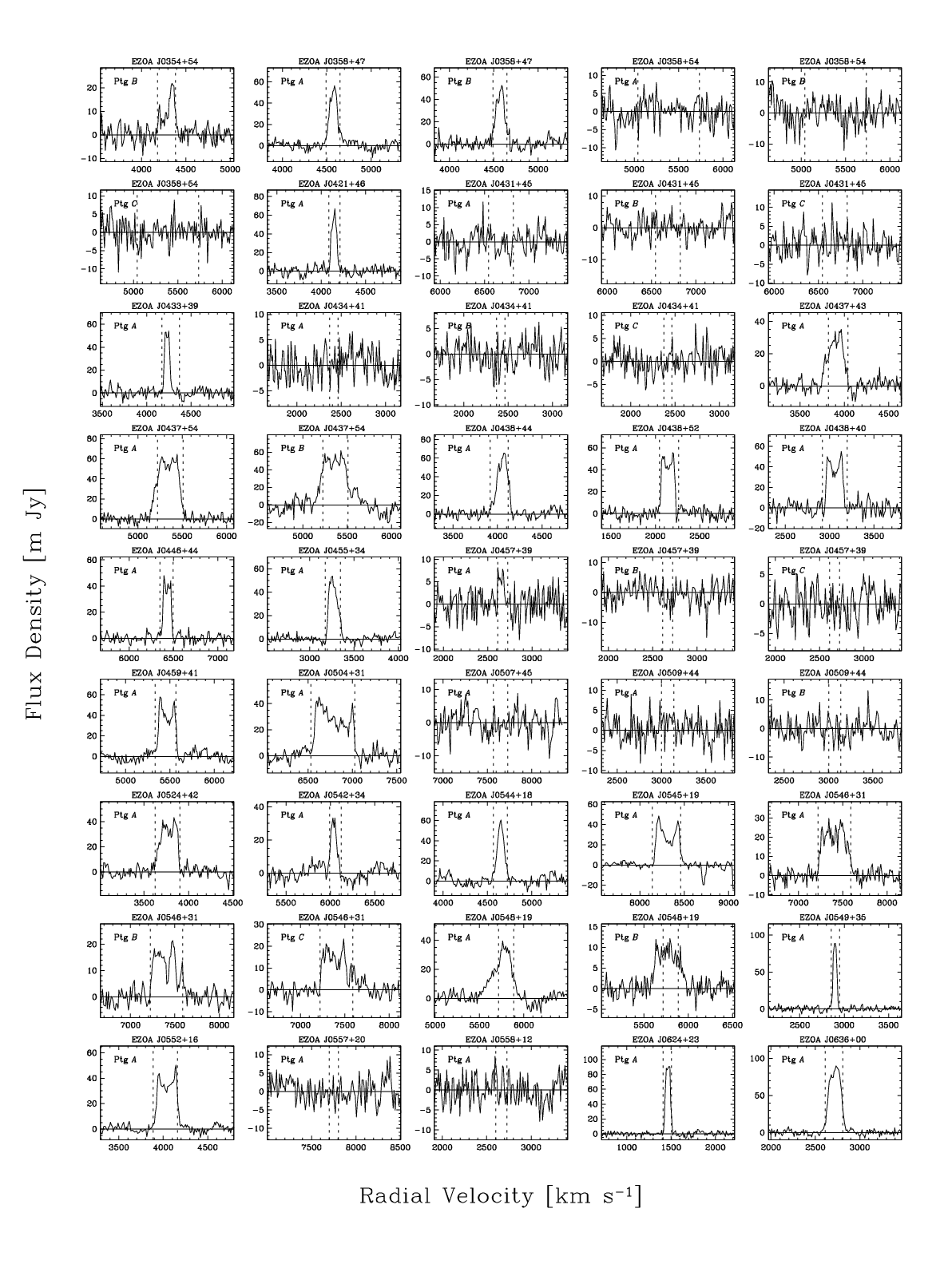}
\vspace{-1cm}
\caption{NRT 21cm \HI\ line spectra -- {continued.}
}
\end{figure*}

\newpage
\clearpage 

\begin{figure*}[h]
\centering
\addtocounter{figure}{-1}
\includegraphics[width=0.98\textwidth]{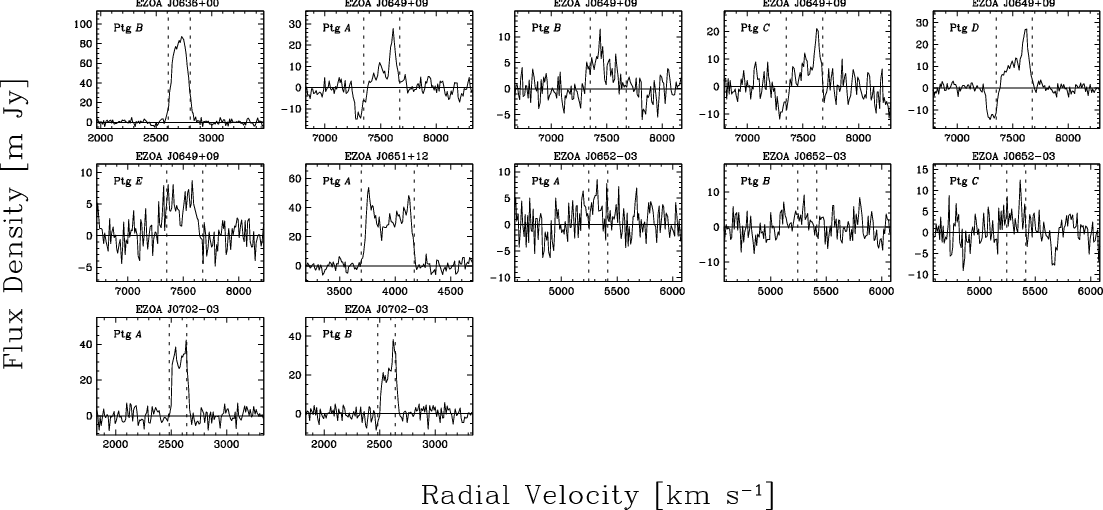}
\caption{NRT 21cm \HI\ line spectra -- {continued.}
}
\end{figure*}

\begin{figure*}[h]
\centering
\includegraphics[width=0.8\textwidth]{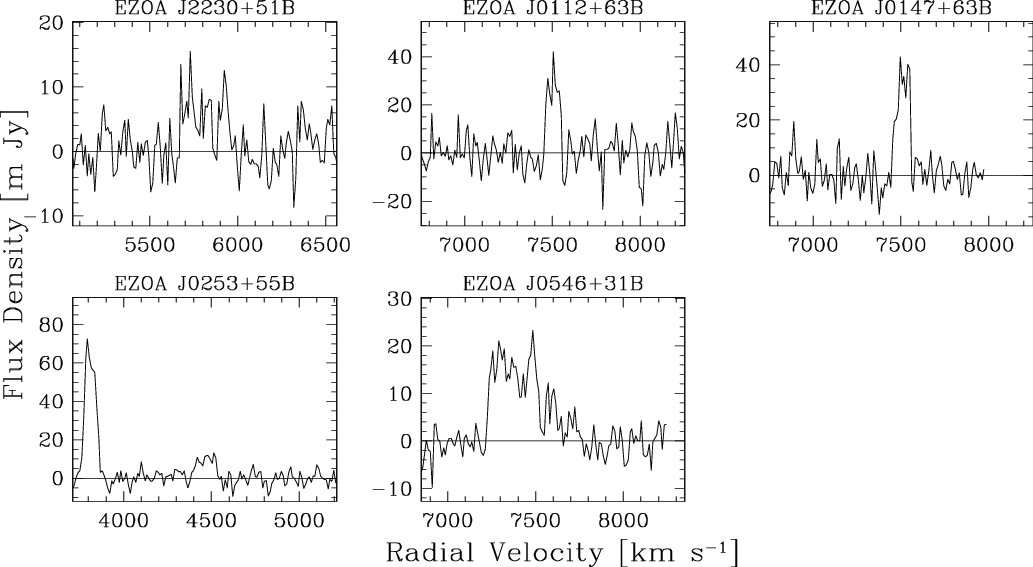}
\caption{Same as Fig.~\ref{figspectra} for serendipitous detections in the
  telescope beam. 
}
\label{fig2spectra}
\end{figure*}

\end{appendix}


\end{document}